\documentclass[review,10pt]{elsarticle}
\usepackage{lineno,hyperref}
\modulolinenumbers[5]
\usepackage{amsmath}
\usepackage{amsfonts}
\usepackage{amssymb}
\usepackage{lmodern}
\usepackage{xfrac}
\usepackage{graphicx}
\graphicspath{ {./figures/} }
\usepackage{subfig}
\usepackage{epsfig}
\usepackage{epstopdf}
\usepackage{textcomp}
\usepackage{multirow}
\usepackage{setspace}
\usepackage{fullpage}
\usepackage{float}
\usepackage{caption}
\usepackage{siunitx}
\usepackage{xcolor}
\usepackage{tabularx, booktabs}
\usepackage{enumitem}
\usepackage{dcolumn}% Align table columns on decimal point
\usepackage{bm}% bold math
\usepackage[T1]{fontenc}
\biboptions{sort&compress}

\begin{document}
\begin{frontmatter}

%\preprint{}

\title{\normalsize{\textbf{Effects of isotherm patterns on cellular interface morphologies of melt pool origin}}}

\cortext[mycorrespondingauthor]{Corresponding author.}
%\cortext[cor1]{\mbox{Corresponding author. \textit{Email address:} \texttt{gsupriyo2004@gmail.com}} (Supriyo Ghosh)}
\author{Saurabh Tiwari}
\author{Supriyo Ghosh\corref{mycorrespondingauthor}}
\ead{supriyo.ghosh@mt.iitr.ac.in; gsupriyo2004@gmail.com}
\address{Metallurgical \& Materials Engineering Department, Indian Institute of Technology Roorkee, Roorkee, UK 247667, India}

%\author[mymainaddress]{\corref{mycorrespondingauthor}}
%\ead{}

%\author{}

%\author[mymainaddress]{\corref{mycorrespondingauthor}}
%\ead{}
%

\date{\today}

\begin{abstract}
Spatiotemporal variation of the thermal gradient in the melt pool inherited from different heat input patterns or other non-equilibrium transient effects during additive manufacturing can significantly affect the resulting subgrain microstructure evolution. To examine the impact of this variation, we approximate the thermal gradient by various isotherm patterns that move with constant velocity following directional solidification. We report the first three-dimensional phase-field simulations to investigate the effects of isotherm patterns on the cellular structures typically observed in solidified melt pools. Results indicate that small variations in the isotherm can considerably impact the microstructural features. We use appropriate statistical characterizations of the solid fraction, solid percolation, and solute partitioning behavior to demonstrate the influence of isotherm patterns on the dendritic structures and semisolid mushy zones. Consistent with experimental observations, we find that non-planar isotherms produce finer cells and reduced microsegregation compared to planar isotherms. Also, we note that a tilt of the isotherm leads to a tilted state of the resulting cellular arrays. Our findings will help in understanding the qualitative aspects of the influence of temperature gradient patterns on the evolution of solidification morphologies, mushy zones, and secondary phases, which are crucial for the macroscopic description of the solidified material.
\end{abstract}

\begin{keyword}
Directional solidification \sep Temperature gradient \sep Isotherm pattern \sep Cellular growth  \sep Phase-field
\end{keyword}

\end{frontmatter}

\section{Introduction}\label{sec_intro}

During the selective laser melting additive manufacturing (AM) process, melting of the feedstock material typically results in a liquid melt pool, which subsequently undergoes a directional solidification process (for recent reviews, see ~\cite{debroy_additive,liu2022additive,haghdadi2021additive_review,review_meltpool,sanchez2021_review}). The resulting solid-liquid interface often adopts a cellular morphology, which in the ideal condition grows with a constant velocity $V$ under thermal gradient $G$. These parameters control the resulting microstructure features, including the grain/subgrain size, growth morphology, and elemental segregation, which determine the properties of the solidified material. Microstructure evolution resulting under AM conditions has been extensively studied through experiments and numerical simulations~\cite{Trevor2017,karayagiz2020,hecht2019am,acharya2017prediction}. For dilute binary alloys, a well-known sequence of morphological transition takes place by continuously decreasing $G$ (or equivalently increasing $V$) from their values for a marginally unstable planar interface: planar $\rightarrow$ cellular $\rightarrow$ dendritic $\rightarrow$ cellular$\rightarrow$ planar. The critical values of $G$ and $V$ required for these transitions are estimated by the constitutional supercooling and absolute stability criteria~\cite{Mullins1964,dantzigbook}.

\begin{figure}
\centering
\includegraphics[width = 0.9\textwidth]{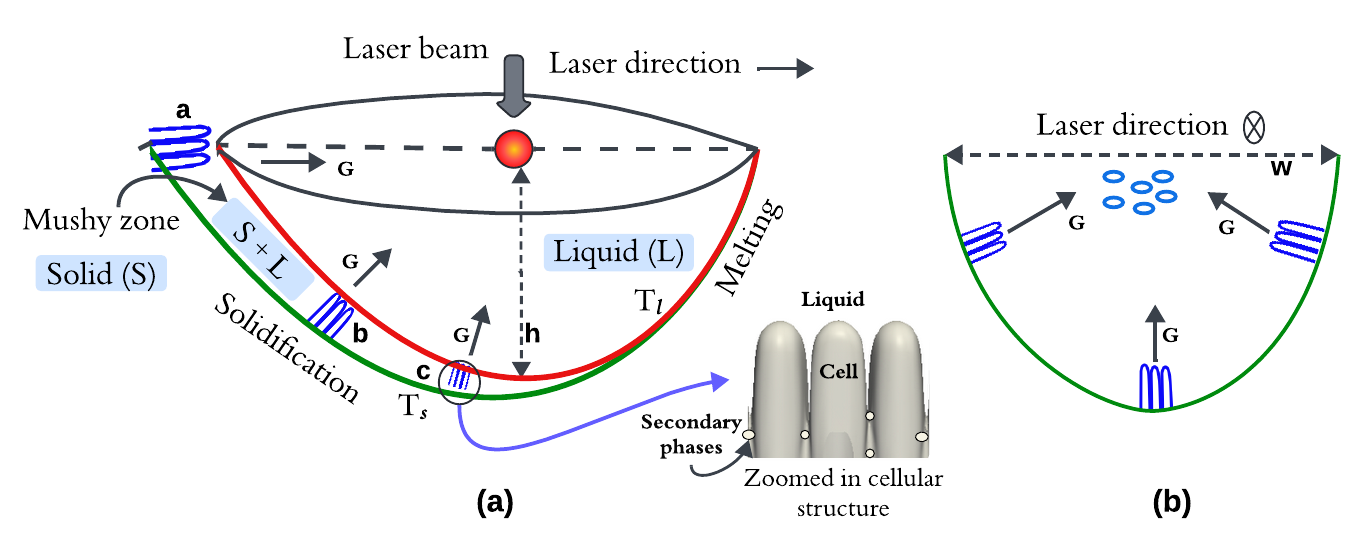}
\caption{Schematic of cellular growth in (a) longitudinal and (b) transverse sections of the melt pool. Solidification and melting fronts are shown. Solidification begins at the melt pool boundary, which can be approximated by the solidus temperature isotherm ($T_s$) taken from the alloy phase diagram, and $T_l$ is the liquidus temperature ahead of the growth front. The two-phase structure (mushy zone) is indicated. The cells (denoted as a, b, c) grow toward the maximum temperature gradient direction, which varies with melt pool location. Thus, a temperature gradient pattern is generated in the liquid ahead of the growth front within the frame of reference of the melt pool. As a result, the cells could grow under different temperature isotherms. A zoomed-in cellular structure details the cellular fronts with segregation-induced secondary phases in the mushy zone. (b) Two different views of cellular fronts (columnar and circular) are shown in a transverse section of the melt pool. Melt pool height ($h$) and width ($w$) are depicted for reference.}
\label{fig_schematic}
\end{figure}

The most common interface morphology that forms during melt pool solidification is the cellular structure constituted by smooth ``fingers'' of solid separated by liquid grooves (Fig.~\ref{fig_schematic}). Therefore, we work with a cellular growth problem at high velocity (near absolute stability). The effects of $G$ and $V$ \textit{via} cooling rates ($GV$) on cellular growth are well understood: an increase of the control parameters leads to finer structures and increased compositional variations (\textit{i.e.}, microsegregation and solute trapping) with a clear trend observed between simulations and experimental counterparts~\cite{Trevor2017,karayagiz2020,hecht2019am,acharya2017prediction}. It is generally known that cellular structures grow epitaxially from the melt pool boundary toward the top center of the melt pool, where has the highest temperature. As such, the largest temperature gradient direction lies along the direction perpendicular to the melt pool boundary. Thus, in the ideal condition, when the crystal grows only by the effect of the thermal gradient, the temperature isotherm ahead of the solidification front remains planar and perpendicular to the growth direction. This is, however, different from the observations reported in the literature~\cite{li2017melt_pulsed,scan_sun2018effect,yao2023melt}, which suggest that other factors including a variation in the direction of $G$ may contribute to the growth behavior. Recent experiments have shown that different patterns of the heat source or other non-equilibrium transient effects lead to microscopic variations of $G$ close to the melt pool boundary, dramatically affecting the resulting cellular morphology~\cite{li2017melt_pulsed,scan_sun2018effect}. Since the shape of the melt pool primarily determines the direction of the thermal gradient, the energy density of the heat source plays a significant role in the resulting temperature curves and, subsequently, the melt pool solidification process~\cite{review_meltpool,yao2023melt}. In experiments, the bottom of the melt pool becomes more curved with high energy density of the heat source, leading to more lateral movement of the solid-liquid interface and, hence, deviation from the ``normal growth'' hypothesis results. During measurements, a global misorientation (\textit{e.g.}, of instrumental origin) and the curvature of the isotherms (\textit{e.g.}, due to intrinsic effects such as the difference in thermal conductivity between solid and liquid) may play a role in modifying the local directions of $G$~\cite{mathis_transverse}. Thus, numerical simulations of AM using a planar isotherm of $G$ may not accurately describe the cellular growth behavior observed in experiments.

Moreover, the traverse path of the laser beam affects the temperature gradient patterns generated in the melt pool. Another approach to modify the thermal field is to modify the laser beam profile. Laser oscillation is also an alternative approach. The resulting solidification patterns depend on the local temperature fields near the growth interface. Thus, a slight deviation of the $G$ profiles through the above means can affect the local microstructural features. The movement of the heat source can further deviate $G$ from its general direction of greatest increase in temperature. As a result, cellular growth direction continuously changes to approach the normal to the solid-liquid interface, even forming unique solidification features~\cite{scan_sun2018effect}. Laser scanning patterns greatly affected the solidification morphology in~\cite{dinda2013}. Significant modifications of the maximum heat flow directions and, thus, solidification morphologies resulted owing to various scan patterns in~\cite{debroy2015texture}.  Modifying the beam shape also affected the $G$ profiles generated in the melt pool, controlling the solidification morphology in~\cite{scan_khairallah2020ca,scan_roehling2020controlling,ebrahimi2023revealing}. Further, customizing the scan strategy between continuous and pulsed beams and further controlling their movement led to spatiotemporal variations in $G$, resulting in refined microstructures and reduced solute segregations~\cite{li2017melt_pulsed}. Therefore, modification of the thermal gradient pattern can provide an alternative pathway for controlling the printability and mechanical properties of the solidified material~\cite{chen2024situ}. 

In view of the above general grounds, it is natural to consider a slight variation in the $G$ patterns to investigate microstructure evolution under AM conditions. As mentioned earlier, cellular structures grow opposite to the heat flow direction, perpendicular to the solid-liquid interface. Thus, when this direction and the temperature gradient are not perfectly aligned, a variety of temperature isotherms is generated in the sample. One of the simplest situations that could be studied is the directional solidification of a planar interface with a small disturbance in the temperature isotherm. As a good starting point in our theoretical analysis, we study the effects of a planar isotherm (\textit{i.e.}, ideal case) disturbed with Gaussian noise and sinusoidal, transverse, curved, and pulsed isotherms on cellular growth. The variation in $G$ within the melt pool due to periodic patterns of the heat source, such as sinusoidal hatching, may induce sinusoidal isotherms in the sample~\cite{sinusoidal_mussatto2022laser}. The pulsed thermal fields represent the variations in heat input during the laser-on and laser-off periods~\cite{li2017melt_pulsed,pulsed_cai2024grain}. Moreover, the geometry of the moving melt pool boundary with its three-dimensional trajectory possesses a curvature (particularly close to the bottom), which makes a genuine physical effect of melt pool disturbance in terms of curved isotherms that could affect the local cellular growth. 

Due to the extremely fine scales of the melt pool, \textit{in situ} measurements of the solidification conditions are challenging. Thus, numerical simulations based on finite element analysis are often used to determine the thermal history in the melt pool. From these simulations, the values of $G$ and $V$ along the melt pool boundary are determined, as detailed in Refs.~\cite{ghosh2017primary,karayagiz2020}. However, these numerical studies do not consider various sources of melt pool ``noise'' as described above, for which a disturbance in $G$ might occur, leading to variations in the resulting cellular patterns. It should be noted that microstructure simulations on typical scales of the melt pool (\textit{e.g.}, width of $\approx$ \SI{150}{\micro\meter} and height $\approx$ \SI{50}{\micro\meter}) are computationally prohibitive, particularly when using high-fidelity methods such as the phase-field~\cite{ghosh2022tusas,ghosh2018single}. Thus, to simplify numerical calculations, the values of $G$ are often taken as constant to study cellular growth over large microscopic volume elements along the melt pool boundary relative to the grain size. However, this may lead to some aleatory uncertainty in the computed growth behavior of the columnar cells compared to experimental measurements~\cite{ghosh2019uncertainty,ghosh2020statistical}.  Therefore, incorporating the effects of isotherm geometry in numerical simulations in terms of various isotherm patterns, such as transverse and curved, could better describe the solidification processes near the respective curvature regions along the melt pool boundary. 

As a case study, we examine the dynamics of cellular substructures generated by the above model isotherms to provide insight into how the presence of a ``disturbance'' in the melt pool influences the solidification patterns generally. While this mechanism is interesting to speculate, its importance on cellular growth has not yet been explored. To our knowledge, there have been limited works on the effects of geometric and thermal variations due to the melt pool disturbance during AM processing. Thus, how a melt pool disturbance acts on local temperature isotherms and, hence, resulting columnar structures remains unclear. To address this, we hypothesize the resulting variations in $G$ by various isotherm patterns assuming directional solidification. Some of these effects in simulations and their experimental counterparts have long been known in other extended out-of-equilibrium systems, including eutectics~\cite{mathis_transverse,kang2024highly,zhang2022_shock}. However, they would be worth studying for non-equilibrium cellular solidification of melt pools. The purpose of this paper is to compute cellular interface morphologies when the temperature isotherms in the melt deviate from the standard planar representations due to small variations in melt pool temperature fields.

The remaining of the article is organized as follows. In the following Sec.~\ref{sec:method}, we briefly describe the phase-field solidification model and the model parameters, including the different forms of temperature isotherms employed in simulations. In Sec.~\ref{sec:results}, we report three-dimensional simulations of binary alloy solidification for the assumed isotherm patterns. In Sec.~\ref{sec:discussion}, we critically discuss the results and compare them with experiments from the literature. We summarize and conclude in Sec.~\ref{sec:summary}. 

\section{Phase-field model}\label{sec:method}
The phase-field method (for reviews, see~\cite{boettinger2002,steinbach2009,chen2024_review}) is the most high-fidelity technique for simulating cellular structures in three dimensions (3D), hence used in this work. We use the model of Echebarria \textit{et al.}~\cite{Echebarria2004}, which solves the temporal evolution of the phase-field ($\phi$) and alloy concentration ($c$) with a frozen temperature ($T$) approximation coupled by their respective equations of motion. These fields vary smoothly within the diffuse interface between their values in the bulk liquid ($\phi$ = -1) and solid ($\phi$ = 1) phases, circumventing the requirement for interface tracking (\textit{i.e.}, -1 < $\phi$ < 1) in simulations. The validation and applications of the model to study AM microstructures can be found in our previous publications~\cite{ghosh2017primary,ghosh2018ti64,ghosh20183d,ghosh2018single,ghosh2022tusas} and elsewhere~\cite{chouhan2024modeling,sahoo2016phase,hecht2019am,farzadi2008,wang2019investigation}. Here, we briefly describe the model for implementing various temperature gradient patterns in the simulation code.

We must admit that the model~\cite{Echebarria2004,karma2001} is based on the following assumptions: (i) melt convection has a negligible effect, so mass transfer is controlled by solute diffusion; (ii) since heat diffuses much faster than solute, the effect of latent heat is ignored by imposing a frozen temperature approximation; (iii) thermophysical properties of the alloy material remain constant; (iv) solute diffusivity in the solid is very small compared to liquid diffusivity, so the solid-state diffusivity is neglected; and (v) the solid-liquid interface obeys local equilibrium. Therefore, the simulation results can be used as a reference for solute diffusion and anti-trapping flux limited cellular growth in the AM solidification regime.

The temporal evolution of $\phi$ is given by
\begin{eqnarray}\label{eq_phi}
\tau_0\, a(\hat{n})^2\,\frac{\partial \phi}{\partial t} = W_{0}^{2} \nabla \cdot \left[{a(\hat{n})}^2 \nabla\phi\right] + W_{0}^{2} \sum_{i=1}^{3} \partial_i \left[a(\hat{n})\, \frac{\partial a(\hat{n})}{\partial(\partial_i \phi)}\, |\nabla\phi |^2 \right]
+ \phi -\phi^3 -  \lambda\, (1-\phi^2)^2\, \left[U + \theta  \right],
 \end{eqnarray}
where $t$ is the time, $\tau_0$ is the relaxation time for $\phi$, $W_0$ is the interface thickness, $\lambda$ is a dimensionless constant, and $a(\hat{n})$ is the crystal anisotropy function. The subscript $i$ in $\partial_i$ denotes partial differentiation in the Cartesian direction $i$, with $i$ =1, 2, and 3 represents the directions $x$, $y$, and $z$ respectively. We assume a four-fold anisotropy given by
\begin{equation}\label{eq_anisotropy}
a(\hat{n}) = 1 - \epsilon_4\, \left[3- 4\,(n_{x}^{4}+n_{y}^{4}+n_{z}^{4})\right],
\end{equation}
where $\hat{n} = \vec{\nabla} \varphi/|\vec{\nabla} \varphi|$ is the unit vector normal to the interface and $\epsilon_4$ is the strength of the anisotropy. The concentration field is expressed as 
\begin{equation}
U = \frac{\exp(u) - 1}{1-k_e},
\end{equation}
where the chemical potential $u$ is given by 
\begin{equation}
u = \ln \left(\frac{2c\,k_e/c_0}{1+k_e-(1-k_e)\,\phi}\right),
\end{equation}
where $c_0$ is the alloy composition and $k_e$ is the equilibrium partition coefficient given by 
\begin{equation}
k_e = \frac{c_s}{c_l},
\end{equation}
where $c_s$ and $c_l$ are the solute concentrations in the solid and liquid sides of the interface, respectively.

We simulate the temperature field $\theta$ (Eq.~\eqref{eq_phi}) with a frozen temperature approximation:
\begin{equation}\label{eq_theta}
\theta = \frac{T-T_0}{\Delta T_0},
\end{equation}
where $T_0$ is the interface temperature and $\Delta T_0$ is the freezing range taken as
\begin{equation}
 \Delta T_0 = \frac{m_l\, c_0\,(1-k_e)}{k_e},
\end{equation}
where $m_l$ is the liquidus slope.
In our approach, the linear field of $T$ moves with directional velocity $V$ under various temperature patterns of magnitude $G$ in the growth direction ($z$-axis). Thus, the temperature isotherms take the following forms:
\begin{equation}\label{eq_planar}
T = T_0 + G\,(z-V\,t), \ \ (\text{planar})
\end{equation}
\begin{equation}\label{eq_noise}
T = T_0 + G\,(z-Vt + A_n\, \delta), \ \ (\text{noise})
\end{equation}
\begin{equation}\label{eq_sinusoidal}
T = T_0 + G\,(z-V\,t + A_s\, \sin(2\, \pi\, f\, x/N_x)), \ \ (\text{sinusoidal})
\end{equation}
\begin{equation}\label{eq_transverse}
T = T_0 + G\,(z-V\,t - x\, \tan\varphi), \ \ (\text{transverse})
\end{equation}
\begin{equation}\label{eq_curvature}
T = T_0 + G\,(z-V\,t - A_c\, x^2), \ \ (\text{parabolic curvature})
\end{equation}
\begin{equation}\label{eq_pulse}
T =
\begin{cases}
    T_0 + G\,(z-V\,t) & t\leq t_p \\
    T_0 & t > t_p \, \,(\text{pulsed-beam}).
\end{cases}
\end{equation}
We choose Eq.~\eqref{eq_planar} to be the reference planar isotherm. In Eq.~\eqref{eq_noise}, we implement a planar interface disturbed with Gaussian noise with $A_n$ the noise amplitude and $\delta$ $\in$ [-1, 1] the random number. In Eq.~\eqref{eq_sinusoidal}, $A_s$ is the amplitude, $f$ is the frequency, and $N_x$ the domain size in the $x$-direction. Note that the $x$-axis is normal to the growth direction. In Eq.~\eqref{eq_transverse}, $\varphi$ is the tilt angle to which the planar isotherms are inclined along the $x$-axis. In Eq.~\eqref{eq_curvature}, $A_c$ is the strength of the curved isotherms, assumed parabolic. In Eq.~\eqref{eq_pulse}, we approximate the pulsed laser effects on the temperature variation by applying a planar isotherm for some residence time $t_p$ before setting it off for a time $t_p$. A schematic of these temperature isotherms on a 2D plane is shown in Fig.~\ref{fig_isotherm}.

For convenience, we use a rescaled temperature field for $\theta$ (Eq.~\eqref{eq_theta}) to be used in Eq.~\eqref{eq_phi} as
\begin{equation}
    \theta = \frac{z - V\,t}{l_T},
\end{equation}
where $l_T$ is the dimensional thermal length given by
\begin{equation}\label{eq_thermal}
l_T = \frac{\Delta T_0}{G}.
\end{equation} 

The temporal evolution of the composition field in terms of $c$ is given by
\begin{equation}\label{eq_c}
\frac{\partial c}{\partial t} = -\nabla \cdot \left[ - \frac{1}{2}\,(1+\phi)\, \tilde{D} \, c \, \exp(u)^{-1}  \, \nabla\exp(u) +\frac{1}{2\sqrt{2}}\, W_0\, (1-k_e)\, \exp(u)\, \frac{\partial \phi}{\partial t}\, \frac{\nabla\phi}{|\nabla \phi|}\right],
\end{equation}
where the first term on the right-hand side represents the standard Fick's diffusion equation, and the last term is the so-called anti-trapping current, which avoids spurious solute concentration effects arising from the use of a large $W_0$ in simulations. We neglect convection in the liquid; thus, the solute is transported only by diffusion. $\tilde{D}$ is the dimensionless liquid diffusion coefficient.

The model Eqs.~\eqref{eq_phi} and~\eqref{eq_c} are solved in dimensionless units using the unit of length scale
\begin{equation}\label{eq_lambda}
W_0 = \frac{8\, d_0\, \lambda}{5\sqrt{2}},
\end{equation}
and the unit of time scale
\begin{equation}\label{eq_tau}
\tau_0 = \frac{0.6267\, \lambda\, W_0^2}{D_l}.
\end{equation}
For example, the liquid diffusion coefficient ($D$) is made dimensionless using
\begin{equation}\label{eq_diffusion}
\tilde{D} = \frac{D\, \tau_0}{W_0^2},
\end{equation}
which sets the dimensionless solutal diffusion length given by 
\begin{equation}
l_D = \frac{2D}{V W_0}.
\end{equation}
Similarly, $l_T$ (Eq.~\eqref{eq_thermal}) can be made dimensionless by $l_T/W_0$. The capillary length $d_0$ in Eq.~\eqref{eq_lambda} is taken as $\Gamma/\Delta T_0$, where $\Gamma$ is the Gibbs-Thomson constant. The $W_0$ and $\tau_0$ are coupled \textit{via} $\lambda$ for the requirement of zero interface kinetics according to the thin-interface limit of the model~\cite{Echebarria2004,karma2001}. 

\subsection{Simulation setup and parameters}\label{sec:parameters}
In this study, we use Ni-Nb as a sample material, assuming a quasi-binary Inconel 718 alloy~\cite{knorovsky1989inconel}. The material parameters are given in Table~\ref{table_param_pf}. Unless otherwise mentioned, simulation parameters or output quantities described below are non-dimensional. We solve model Eqs.~\eqref{eq_phi} and~\eqref{eq_c}, on a $N_x \times N_y \times N_z$ domain  of 128 $\Delta x$ $\times$ 128 $\Delta y$ $\times$ 1800 $\Delta z$. We apply a uniform mesh spacing $\Delta x/W_0 = \Delta y/W_0 = \Delta z/W_0$ = 0.8, time step $\Delta t/\tau_0$ = 0.05. We use $W_0$ = $10^{-8}$ m, leading to $\lambda$ = 1.4 (Eq.~\eqref{eq_lambda}) and $\tau_0 = 3.6 \times 10^{-9}$ s (Eq.~\eqref{eq_tau}). Lower values of $W_0$ and $\tau_0$ are required to resolve the scales in AM microstructures, making phase-field computations very costly. We use an explicit Euler finite difference algorithm, with a standard nine-point stencil Laplacian operator and no-flux boundary conditions for both $\phi$ and $c$ in all directions. A comprehensive description of numerical schemes and algorithmic descriptions can be found in~\cite{provatasbook,Echebarria2004}.

\begin{figure}[htbp]
\centering
\includegraphics[width=\textwidth]{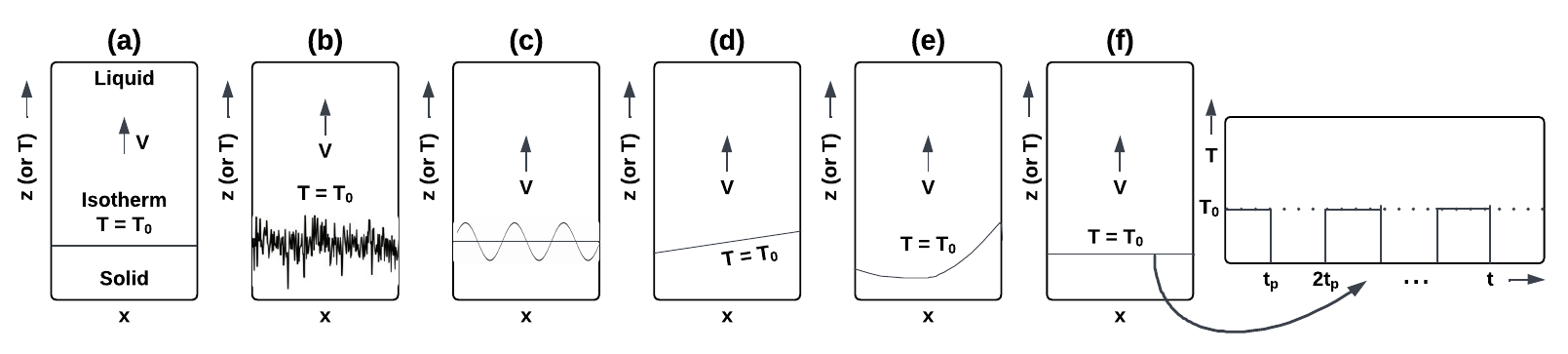}
\caption{Two-dimensional view of the isosurface at isotherm $T = T_0$ (Eq.~\eqref{eq_theta}). We assume various isotherm patterns, including (a) planar, (b) planar with Gaussian noise, (c) sinusoidal, (d) transverse, (e) curved, and (f) pulsed to simulate upward directional solidification. As a first guess, the pulsed beam-induced isotherm is approximated by the effects of varying heat input: the laser-on period with a planar isotherm and the laser-off period with a uniform melt temperature. During directional solidification, these isotherms are characterized by a constant thermal gradient $G$, which moves upward with velocity $V$ in the direction ($z$) from solid to liquid.}\label{fig_isotherm}
\end{figure}

The growth direction is $z$; thus, $x-y$ planes become normal to the growth direction. Simulations begin with a thin layer of solid (of height $20\ \Delta z$) from the bottom of the simulation box, with $\phi$ set as $\phi (z,\ t = 0) = -\tanh((z-20\ \Delta z)/\sqrt{2})$ and initial Nb concentration of $k_ec_0$ in the solid and $c_0$ in the liquid. The initial interface position corresponds to the liquidus temperature for the concentration $c_0$; \textit{i.e.}, $T = T_0$ in Eq.~\eqref{eq_theta}, as illustrated in Fig.~\ref{fig_isotherm}. A random noise of amplitude $\pm$ 0.01 is added to the initial configurations to trigger cellular growth. A maximum simulation runtime of $80\,000\, \Delta t$ is used to compare the simulation results with various isotherm patterns in the steady state. 
 
Since we aim to study the influence of isotherm patterns on cellular structures, the number of relevant parameters can be further reduced. We use a constant value of $G$ = $10^7$ K m$^{-1}$ and $V$ = 0.1 m s$^{-1}$ typical for laser powder bed fusion simulations~\cite{ghosh2023_review}, leading to $l_D$ = \SI{0.06}{\micro\meter} and $l_T$ = \SI{11.38}{\micro\meter}. Thus, we work in the limit of high $G$ and $V$, \textit{i.e.}, $d_0 << l_D << l_T$. These ranges are typical for directional solidification experiments with metal alloys. Moreover, $l_T >> l_D$ signifies the solidification regime far above the constitutional supercooling limit, given by $l_T$ = $l_D$, for the onset of cellular growth.

Further, in the isotherm equations, we use $A_n = 0.5$ (Eq.~\eqref{eq_noise}), $A_s = 0.5$ (Eq.~\eqref{eq_sinusoidal}), $\varphi$ = 15$^{\circ}$ (Eq.~\eqref{eq_transverse}), and $A_c = 0.0005$ (Eq.~\eqref{eq_curvature}). These are just reference values for a comparative study of the effects of using various isotherms (isosurfaces in 3D) on the resulting cellular structures.  

The size of the simulation box in the growth direction ($N_z$) is taken as $\approx$ \SI{15}{\micro\meter}, which is $\approx$ 250 times $l_D$. The lateral size of the simulation box ($N_x$ or $N_y$) is taken as \SI{1.024}{\micro\meter}, that is, $\approx$ 20$\lambda_c$, where $\lambda_c \approx \sqrt{d_0\,l_D}$ is the characteristic length scale of the cellular pattern arise out of the simulations. These $\lambda_c$ is small enough to capture the well-developed cellular-intercellular structures and the mushy zones between solid and liquid in our simulations. We use our in-house, C programming language-based, parallel phase-field code framework for numerical simulations. The parallel calculations are carried out on a server utilizing eight cores in an Intel Xeon Silver 4309Y CPU operating at 2.8 GHz supported by 32 GB of RAM. On average, the execution time was $\approx$6 days for $80\,000$ time steps.

\addtolength{\tabcolsep}{2em}
\begin{table}[h]
\centering
\begin{tabular}{l l}
\hline
Initial alloy concentration, $c_0$ 	&	\SI{5}{wt \%} \\
Equilibrium partition coefficient, $k_e$	&	0.48 (dimensionless)	\\
Liquidus slope, $m_l$	&	-10.5 K \%$^{-1}$	\\
%Equilibrium Freezing Range, $\Delta T_0$ = $T_l - T_s$ 		&	57 K	\\
Liquid diffusion coefficient, $D$	& $3 \times 10^{-9}$ m$^2$ s$^{-1}$ \\
%Solid Diffusion Coefficient, $D_s$	& $10^{-12}$ m$^2$ s$^{-1}$	\\
Anisotropy strength, $\epsilon_4$	& \SI{3}{\%}	\\
Capillary length, $d_0$ & $6.4 \times 10^{-9}$ m \\
Gibbs-Thomson coefficient, $\Gamma$ &  $3.65 \times 10^{-7}$ K m \\
Solidification rate, $V$ & 0.1 m s$^{-1}$ \\
Thermal gradient, $G$ & $10^7$ K m$^{-1}$\\
\hline
\end{tabular}
\caption{Material and solidification parameters used in simulations, after Refs.~\cite{nie_2014,knorovsky1989inconel}.}\label{table_param_pf}
\end{table}

\section{Results and comparative analysis}\label{sec:results}

\subsection{Cellular growth in three-dimensions}
For all the isotherm patterns that we have used, the overall dynamics of the breakdown of initially planar interfaces into cellular structures ($\gamma$ phase in Ni-Nb phase diagram~\cite{knorovsky1989inconel,asmdatabase}) is found to be similar. Due to the Mullins-Sekarka instability~\cite{Mullins1964}, perturbations of the initial interface caused by noise grow with time along the $z$-axis. After the initial transient involving submerging, overgrowing, and splitting phenomena of the neighboring cells, their tips reach a steady state, where the large-scale geometrical features (\textit{e.g.}, cell spacings, tip positions) and tip characteristics (\textit{i.e.}, velocity, composition, and temperature) on average do not vary with time at the end of the run. We extract these steady state microstructure signatures for data analysis (Sec.~\ref{sec_analysis}). 

As the cells grow by rapid freezing, the solute is rejected into the liquid around its base (\textit{i.e.}, intercellular grooves), leading to compositional variation across the interfacial region (\textit{i.e.}, microsegregation). The grooves between cells get narrower with increasing depth (from the tip) and eventually coalesce, separating the solute-rich liquid from the groove bottoms as spherical ``droplets''. This process is analogous to the Plateau-Rayleigh instability of a liquid jet and hence related to the Balling phenomena observed in AM experiments~\cite{debroy_additive,liu2022additive,haghdadi2021additive_review,review_meltpool}. Subsequently, the liquid droplets are included in the semisolid mushy zone, while the cell tips and grooves move globally upward. The droplets freeze only at a later time and thus could influence the formation of secondary phases in the subsequent evolution~\cite{tucho_2017,kuo2017effect,kouraytem2021,sun2022hottearing}. The amount of residual liquid decreases with the depth, leading to a transition from liquid-like to solid-like behavior during mushy zone evolution.

\begin{figure}[ht]
\centering
\subfloat[]{\includegraphics[width=0.3\textwidth]{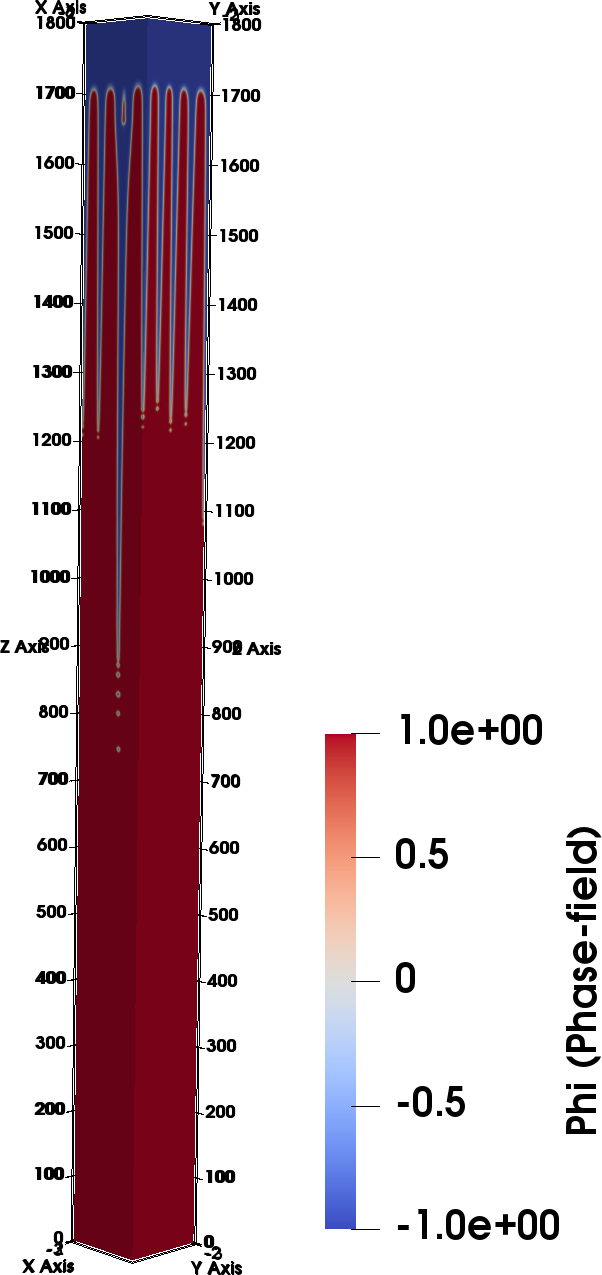}}\hspace{5mm}
\subfloat[]{\includegraphics[width=0.3\textwidth]{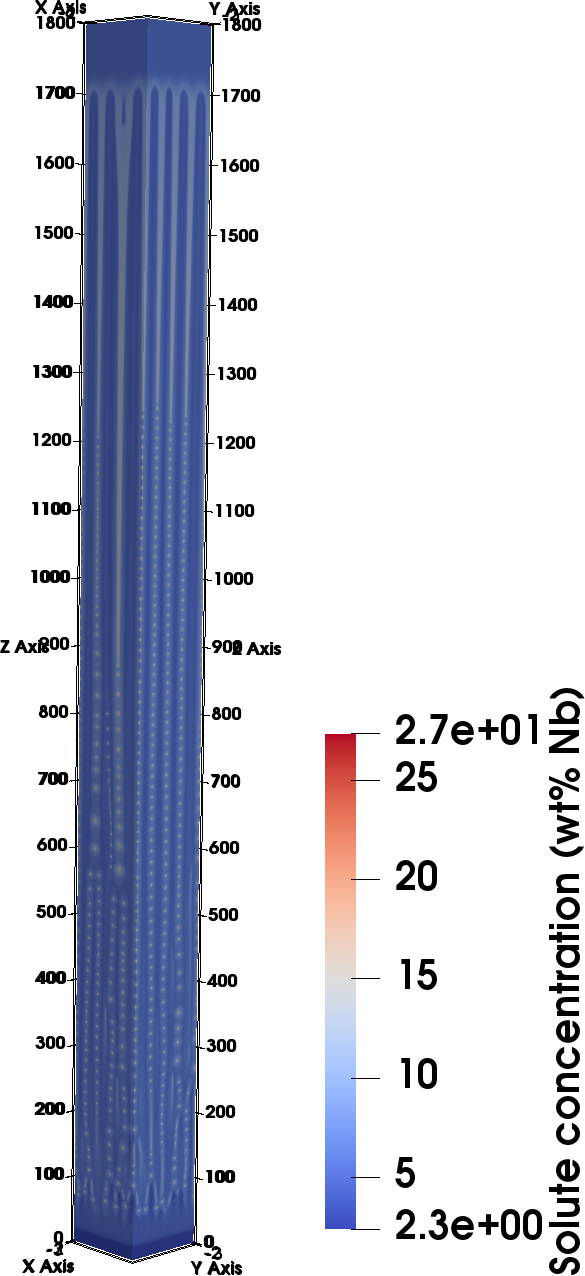}}\hspace{5mm}
\subfloat[]{\includegraphics[width=0.3\textwidth]{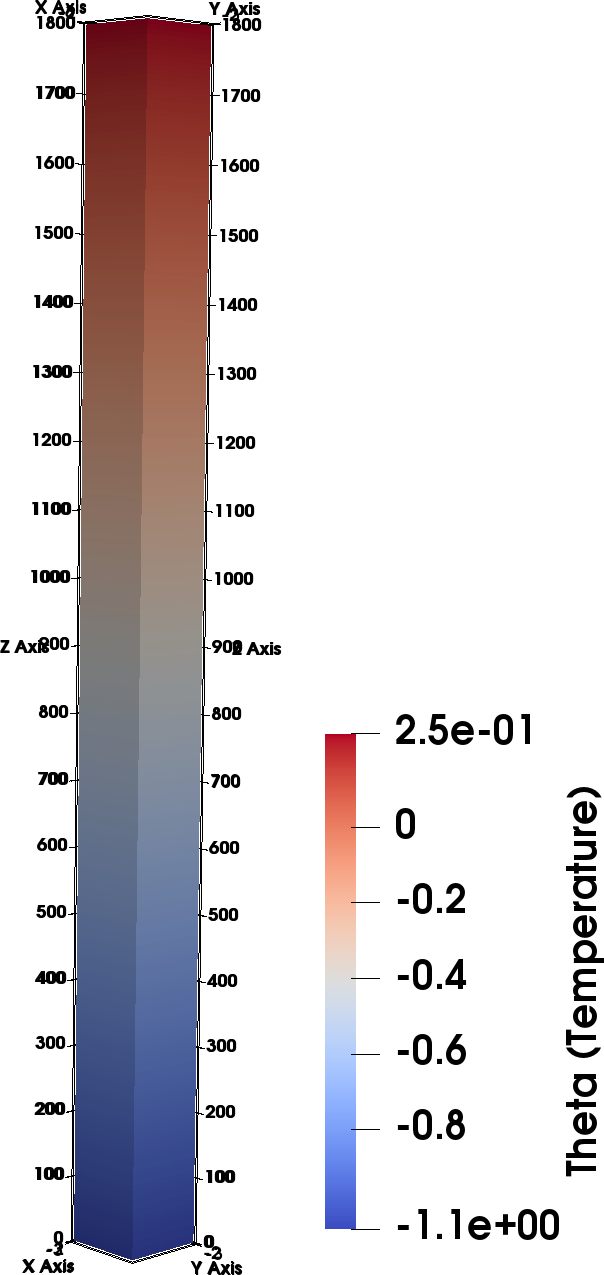}}\hspace{5mm}
%\subfloat[]{\includegraphics[scale=0.39]{3d1_temp}}
\caption{(Color online) Typical 3D representations of the (a) phase-field (Eq.~\eqref{eq_phi}), (b) concentration field (Eq.~\eqref{eq_c}), and (c) temperature field ($\theta$ in Eq.~\eqref{eq_theta}) are shown. The reference isotherm pattern is planar (Fig.~\ref{fig_isotherm}a). The $z$-axis is the growth direction.}\label{fig_cellular}
\end{figure}

Typical 3D simulations of cellular microstructure modeled with the planar isotherm (Eq.~\eqref{eq_planar}) are shown in Fig~\ref{fig_cellular}. The spatial distribution of the phase-field (Fig.~\ref{fig_cellular}a), concentration profile (Fig.~\ref{fig_cellular}b), and temperature field (Fig.~\ref{fig_cellular}c) is plotted. Figure~\ref{fig_cellular}a shows the typical cellular structure with well-developed cell tips, intercellular grooves, and droplets close to the groove bottoms. Figure~\ref{fig_cellular}b shows the concentration profile across the cellular-intercellular regions, with a variance in concentration between 2.3 wt\% Nb and 27 wt\% Nb can be observed. Due to the rapid freezing, the solute does not have sufficient time required for equilibrium partitioning at the solid-liquid interface, leading to a cell core concentration $c<c_0$. This is referred to as the solute trapping. At the same time, the intercellular liquid becomes enriched with solute $c > c_0$. The emission of this solute-rich liquid as spherical ``droplets'' from the grooves between the cells can be visualized in Fig.~\ref{fig_cellular}b. In the context of a Ni-Nb alloy (\textit{i.e.}, quasi-binary Inconel 718), the solute-rich particles could transform into secondary solid phases during the last stage of solidification~\cite{tucho_2017,kuo2017effect,kouraytem2021,sun2022hottearing}.

In Fig.~\ref{fig_cellular}c, we show the trajectory of the temperature field, which is linear and moves with velocity $V$ under a temperature gradient of magnitude $G$ along $z$ so that the interface approximately remains at $T = T_0$ (Eq.~\eqref{eq_theta}). The magnitude of the resulting temperature variation (\textit{i.e.}, undercooling) varies between 63 K (-1.1 $\theta$) close to the bottom and 14 K (0.25 $\theta$) close to the top of the simulation box. For all the isotherms employed, the trends in the temperature field remain similar. This is not surprising in view of the frozen temperature approximation for which interface shape is independent of $T$ and, hence, a function of space and time only. Hence, we do not repeat the visualization for other isotherms here. It should be noted that a frozen temperature field is the only reasonable way to vary the temperature gradient pattern, while keeping the control parameters $G$ and $V$ fixed.

\begin{figure}
\centering
\includegraphics[width=\textwidth]{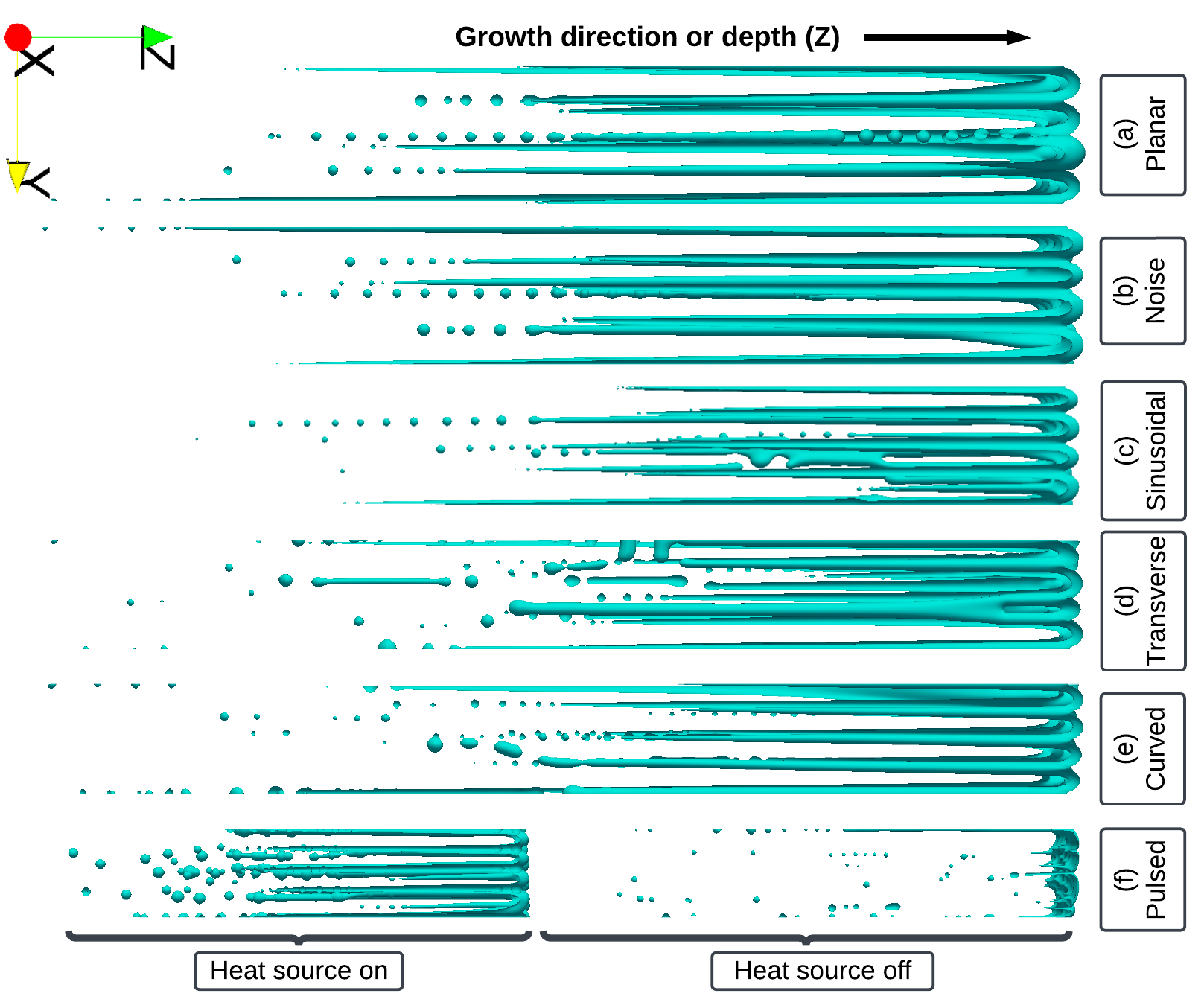}
\caption{(Color online) The spatial representation of the final solid-liquid interface along $z$ in the $y-z$ plane (side view) is plotted for different isotherm patterns. The ``skeleton'' of the cellular structures, with cell tips, intercellular grooves, and the emission of droplets from the groove bottoms can be seen. The phase-field contour of $\phi$ = 0 is used to extract the interface.}\label{fig_side}
\end{figure}

For a comparison of the general effects of various isotherm patterns, we have performed a survey of the phase-field representation of the solid-liquid interface ($\phi$ = 0) from different steady states obtained with various isotherm patterns (Figs.~\ref{fig_side}a-f). On average, the isotherms produce deep cellular shapes separated by long liquid channels with a trail of spherical (and sometimes tubular) droplets emanating from the base region of the cell grooves. The amount and emission frequency of the droplets are apparently different with different isotherms. Specifically, in the pulsed isotherm pattern, we impose the planar isotherm to approximate the laser-on period and assume the thermal gradient in the liquid to be zero during the laser-off period. Thus, cellular structures develop in two stages (Fig.~\ref{fig_side}f). The laser-on period (bottom) produces deep cellular shapes in the same way they develop with other isotherms (Fig.~\ref{fig_side}a); however, the laser-off period (top) produces shallow cells, where the tips are very flat and the grooves are thin with significantly less number of droplets produced from their bottoms. In experiments, a relatively uniform heat transfer resulted along the melt pool solidification front during the laser-off period, resulting in less location-dependent cooling rates~\cite{li2017melt_pulsed}. As a first guess, we approximate the laser-off period with a uniform melt temperature. However, complex temperature gradient patterns can be approximated to model the thermal field generated by the pulse pattern, for example, imposing a decreasing function of $G$ with time, leading to local variations in solidification features within the solidified material (see~\ref{sec_appendix1}).

\subsection{Data analysis: solidification pathways}\label{sec_analysis}
Given the complexity of the microstructure evolution and solidification paths in 3D (Figs.~\ref{fig_cellular} and~\ref{fig_side}), it is certainly challenging to interpret and thus quantify the simulation results. Therefore, we extract the trajectory of the solid and liquid phases from the transverse $x-y$ cross-sections of the 3D snapshots for a comparative analysis of the cellular structures and mushy zones generated by various isotherms (Fig.~\ref{fig_isotherm}).

\begin{figure}[htbp]
\centering
\includegraphics[width=\textwidth]{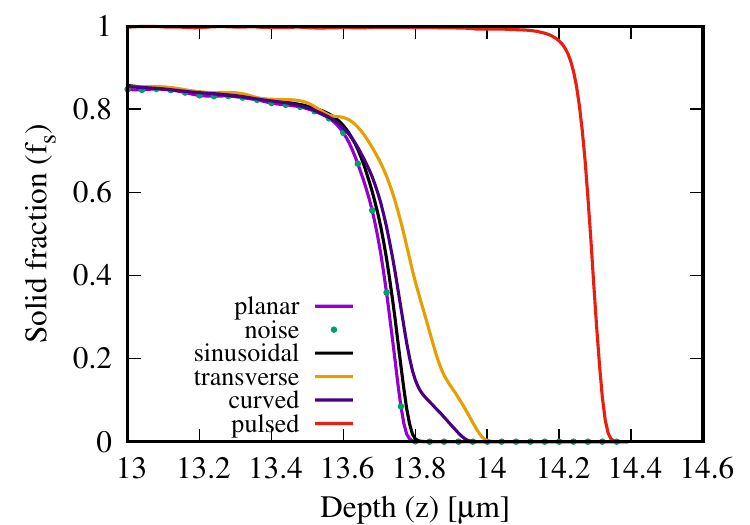}
\caption{(Color online) The variation of solid fraction ($f_s$) between solid ($f_s = 1$) and liquid ($f_s = 0$) in the mushy zone is plotted as a function of depth ($z$). We only present the data over a certain depth of interest to emphasize the variation between different curves. The curves for the planar isotherm and for the case when planar isotherm is disturbed with Gaussian noise show no difference. On average, non-planar isotherms produce a smaller slope or steepness ($df_s(z)/dz$) than planar isotherms.}\label{fig_fraction}
\end{figure}

First, we estimate the area fraction of the solid phase ($f_s$) in each $x$-$y$ plane from the bottom ($z = 0$) to the top ($z = 1800\ \Delta z$) of the simulation box. We use
\begin{equation}\label{eq_solidfraction}
f_s(z\ (\text{or}\, T)) = \frac{1}{N_x\, N_y} \int \frac{1}{2}\, \left[ 1 + \phi(x,\, y,\, z) \right]\, dx\, dy,
\end{equation}
for calculation (where $\phi$ = 1 for solid and $\phi$ = -1 for liquid). It should be noted that the temperature $T$ is proportional to the distance in the growth direction $z$. We show the variation of $f_s$ as a function of $z$ in Fig.~\ref{fig_fraction}. Clearly, $f_s$ = 0 in the liquid ahead of the cell tips and continuously increases behind the tips until $f_s$ = 1 in the solid. Thus, the solid-liquid interface close to the cell tip is roughly given by the $f_s$ curves near $f_s = 0$. Moreover, the steepness of the $f_s$ curves approximated by $df_s/dz$ changes with varying isotherm patterns, indicating variations in groove depths. The $df_s/dz$ varies between 6.3 \SI{}{\micro\meter}$^{-1}$ (planar isotherm) and 2.3 \SI{}{\micro\meter}$^{-1}$ (transverse isotherm). The extension of the mushy zone represented by $0< f_s <1$ thus varies with $df_s/dz$. Comparing the average effects of the isotherm patterns, we find that the length of the mushy zone is smallest with the planar isotherm, while the transverse isotherm produces the longest mushy zone. The longer the mushy zone, the region close to the root of the grooves is ever further
enriched with the solute, and, hence, a much lower temperature is required to freeze the last remaining liquid. In experiments, solidification defects such as microporosity often result when the limited residual liquid in the mushy zone cannot compensate for solidification shrinkage~\cite{sun2022hottearing}. Thus, a longer mushy zone could increase the probability of defect formation in the solidified material. 

\begin{figure}[htbp]
\centering
\includegraphics[width=\textwidth]{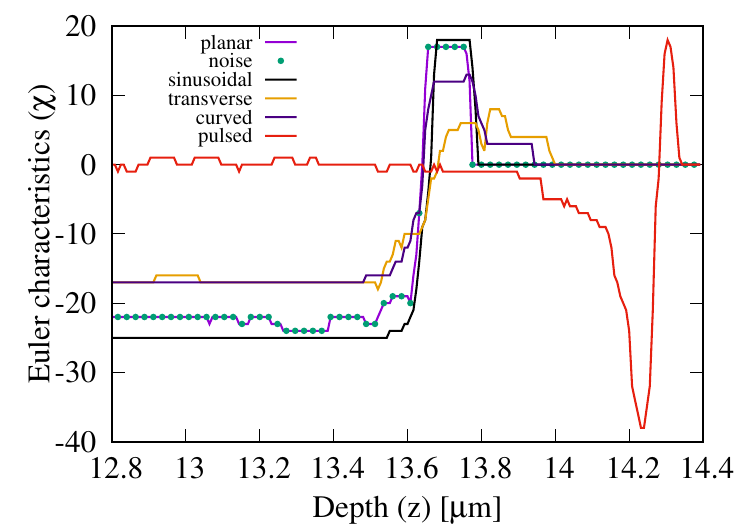}
\caption{(Color online) We present the variation of Euler characteristic $\chi$ as a function of depth to represent the connectivity of the solid and liquid phases in the mushy zone. We only present $\chi$ over a certain depth to highlight the variation between different curves. Solid cells and liquid channels coalesce in the mushy zone. Thus, a larger number of solid features than the liquid features in a plane contributes to +1 to $\chi$, while a larger number of liquid features than the solid features in a plane contributes to -1 to $\chi$. The $\chi$ = 0 represents the ``bridging'' plane, where the number of solid and liquid features is the same. The curves for the planar isotherms with and without the noise show no difference. The maximum in $\chi$ roughly represents the number of cells in the cellular structure.}\label{fig_connection}
\end{figure}

The $f_s$ increases behind the cell tips as a function of decreasing $z$ (increasing depth) until $f_s$ becomes one during terminal solidification (Fig.~\ref{fig_fraction}). The trajectory of $f_s$ is noisy due to the rapid, random merging (connection) and splitting (disconnection) of cellular solid and liquid phases at different depths within the mushy zone. We analyze such complex distribution of solid and liquid phases as a function of $z$ employing the Hoshen-Kopelman algorithm~\cite{hoshen1976percolation} commonly used for the analysis of connected domains in applications such as cluster labeling in percolation phenomena~\cite{percolation_book}. We estimate the Euler characteristic ($\chi$) in terms of $\phi$ in $x$-$y$ planes by 
\begin{equation}\label{eq_euler}
\chi (z) = N_{\phi_{+1}} - N_{\phi_{-1}},
\end{equation}
where $N$ is the count of connected and disconnected domains of $\phi$ = 1 (solid) or $\phi$ = -1 (liquid). Thus, $\chi$ represents the processes of cell tip splitting and merging, which are translated into the connectivity between the solid and liquid phases in a plane. The analysis is shown in Fig.~\ref{fig_connection}. With the solid becoming more connected with increasing depth in the mushy zone, the value of $\chi$ changes from positive to negative, with $\chi$ = 0 signifies a characteristics plane in the mushy zone (so-called ``bridging'' plane), where solid and liquid are very connected. Therefore, the zero crossing of $\chi(z)$ becomes a good measure of the mushy zone depth at which solid percolates in the $x$-$y$ plane. In addition, a section through the cell tips roughly gives the number of individual solid cells in the liquid, representing the maximum in $\chi$. In this context, the cell spacing in 3D is generally measured as~\cite{spacing_3d}
\begin{equation}\label{eq_spacing}
    \lambda_c \approx \sqrt{\frac{A}{n}},
\end{equation}
where $n$ is the number of cells within area $A$, or more accurately by examining the power spectrum, $s_k = |h_k|^2$, of the solid-liquid interface as~\cite{ghosh20183d}
\begin{equation}\label{eq_power}
    \lambda_c \approx 2\pi \frac{\sum s_k}{\sum ks_k},
\end{equation}
where $k$ is a positive wave number. The maximum of $\chi$ curves indicates the variations in cell spacing with different isotherm patterns. Also, the statistics of $\chi(z)$ in the mushy zone are noisy due to the apparent randomness in the coalescence behavior between solid cells, liquid grooves, and liquid droplets in a plane. 

\begin{figure}[ht]
\centering
\includegraphics[width=0.6\textwidth]{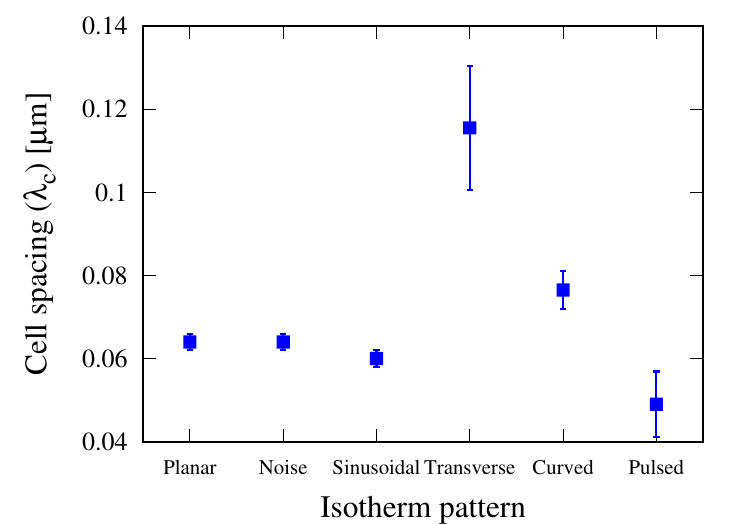}
\caption{The cell spacing ($\lambda_c$) is plotted for various isotherm patterns. Each $\lambda_c$ is plotted with a confidence interval, representing the standard deviation around the mean obtained by averaging the data. On average, the sinusoidal and pulsed isotherm profiles generate finer cells, whereas the transverse isotherm produces the coarsest cellular structure. We anticipate that the differences between $\lambda_c$ values will be more pronounced when using a large simulation box size for which more cells will form, although such work would be computationally intensive.}
\label{fig_spacing}
\end{figure}

Figure~\ref{fig_connection} shows the $\chi(z)$ curves for various isotherm patterns, signifying different cell spacings and mushy zone lengths generated by various isotherm patterns. In particular, the transverse isotherm produces the longest (similar to Fig.~\ref{fig_fraction}), and the pulsed isotherm produces the smallest mushy zone. Also, the number of cells appears to be the highest with the pulsed and sinusoidal isotherms and the smallest with the transverse isotherm. Following Eq.~\eqref{eq_spacing}, the average cell spacing varies between \SI{0.06}{\micro\meter} (sinusoidal and pulsed) to \SI{0.13}{\micro\meter} (transverse), as shown in Fig.~\ref{fig_spacing}. Identical $\lambda_c$ values (within numerical uncertainty) are obtained following Eq.~\eqref{eq_power}. As expected, the planar isotherms with and without the noise produce identical $\chi$ curves or mushy zone evolution. Moreover, the bridging plane characterized by the zero crossing of $\chi(z)$ differs by several diffusion lengths for different isotherms, signifying that the transition from liquid-like (\textit{i.e.}, liquid containing isolated cells) to solid-like (\textit{i.e.}, solid containing isolated liquid droplets) behavior of the mushy zone depends on the isotherm patterns. Such transition during mushy zone evolution could influence the formation of solidification defects such as hot cracking in the solidified material~\cite{sun2022hottearing,kou2015criterion}. 

\begin{figure}[ht]
\centering
\subfloat[Planar]{\includegraphics[width=0.15\textwidth]{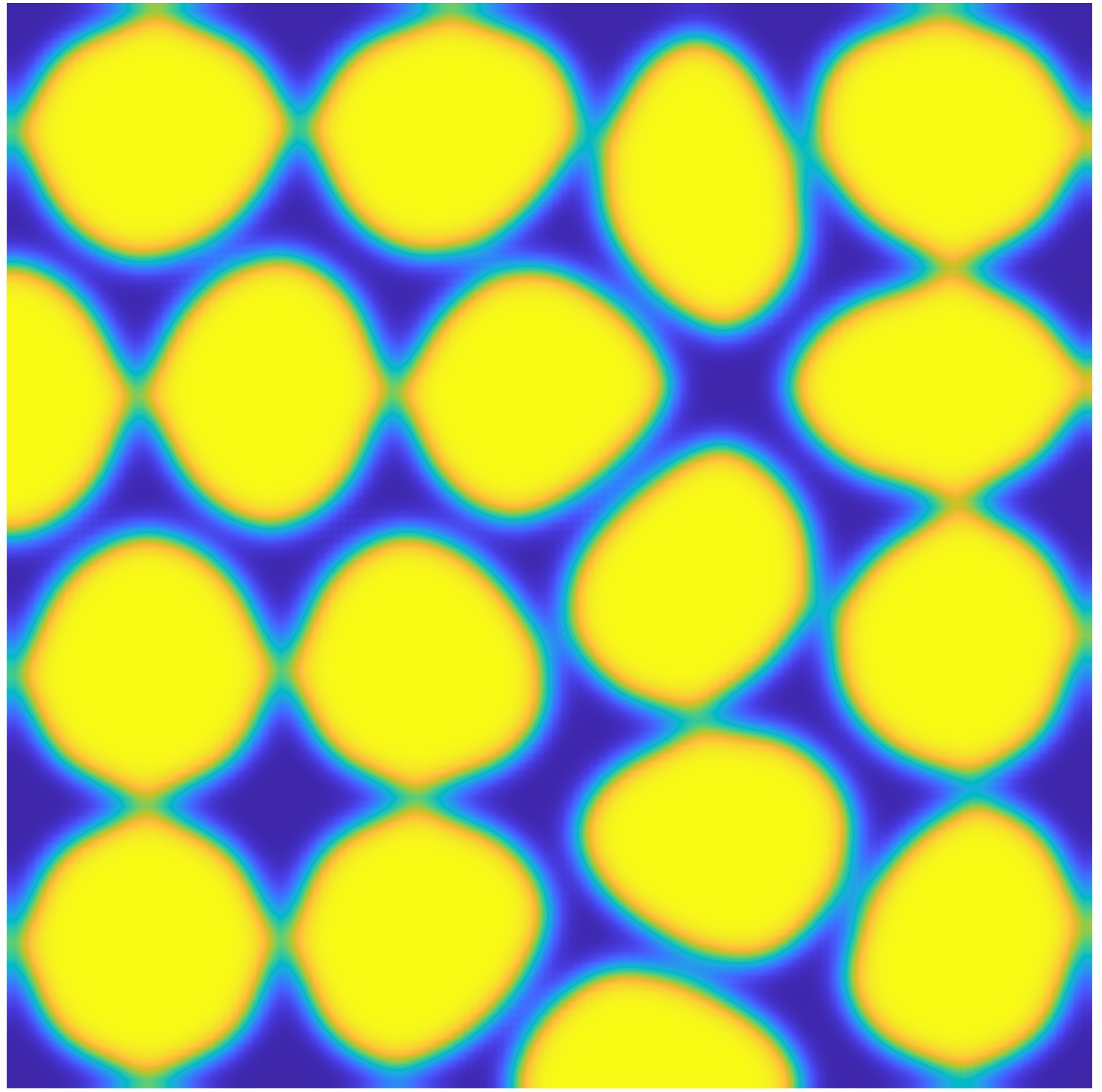}}\hfill
\subfloat[Noise]{\includegraphics[width=0.15\textwidth]{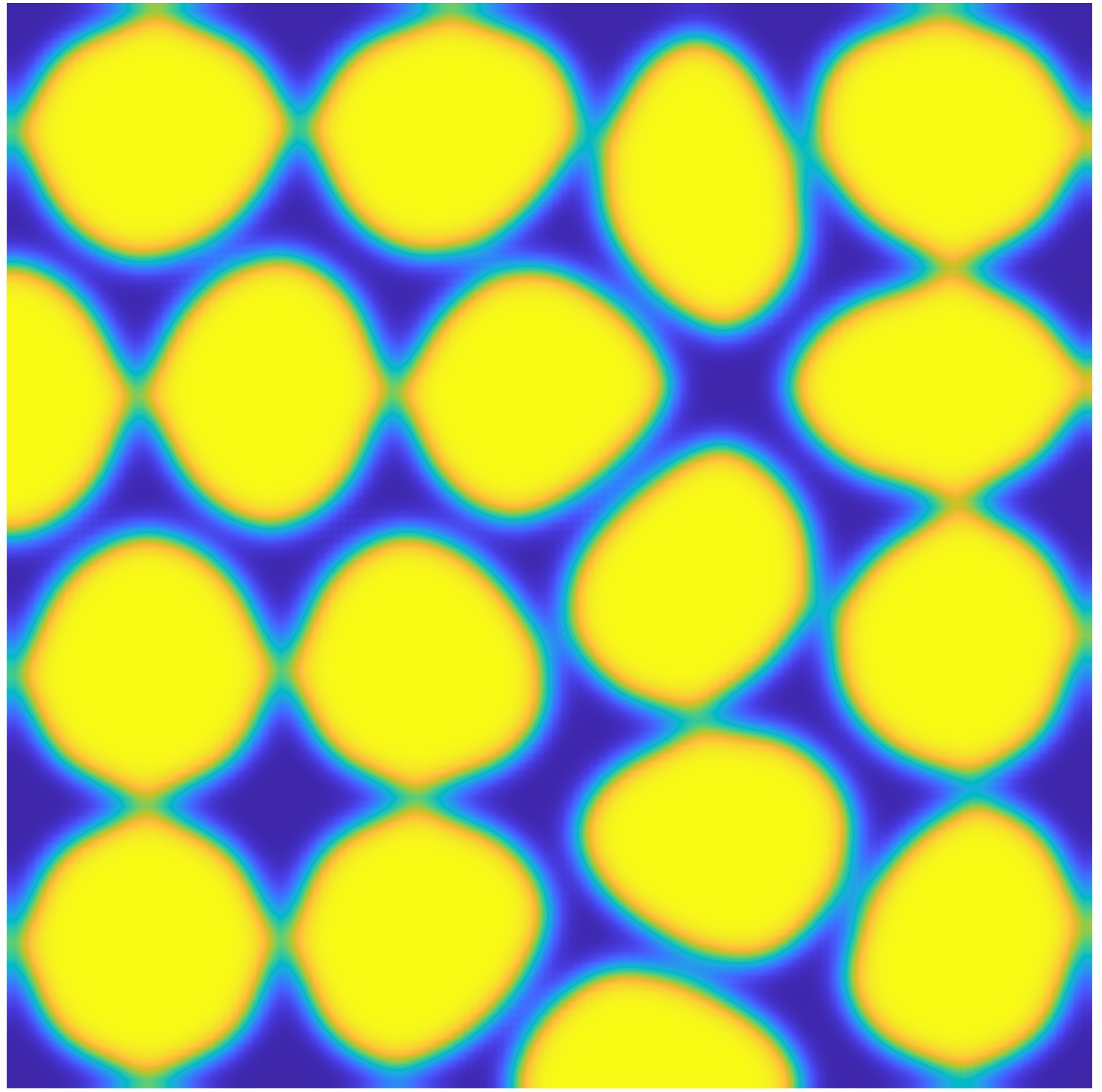}}\hfill
\subfloat[Sinusoidal]{\includegraphics[width=0.15\textwidth]{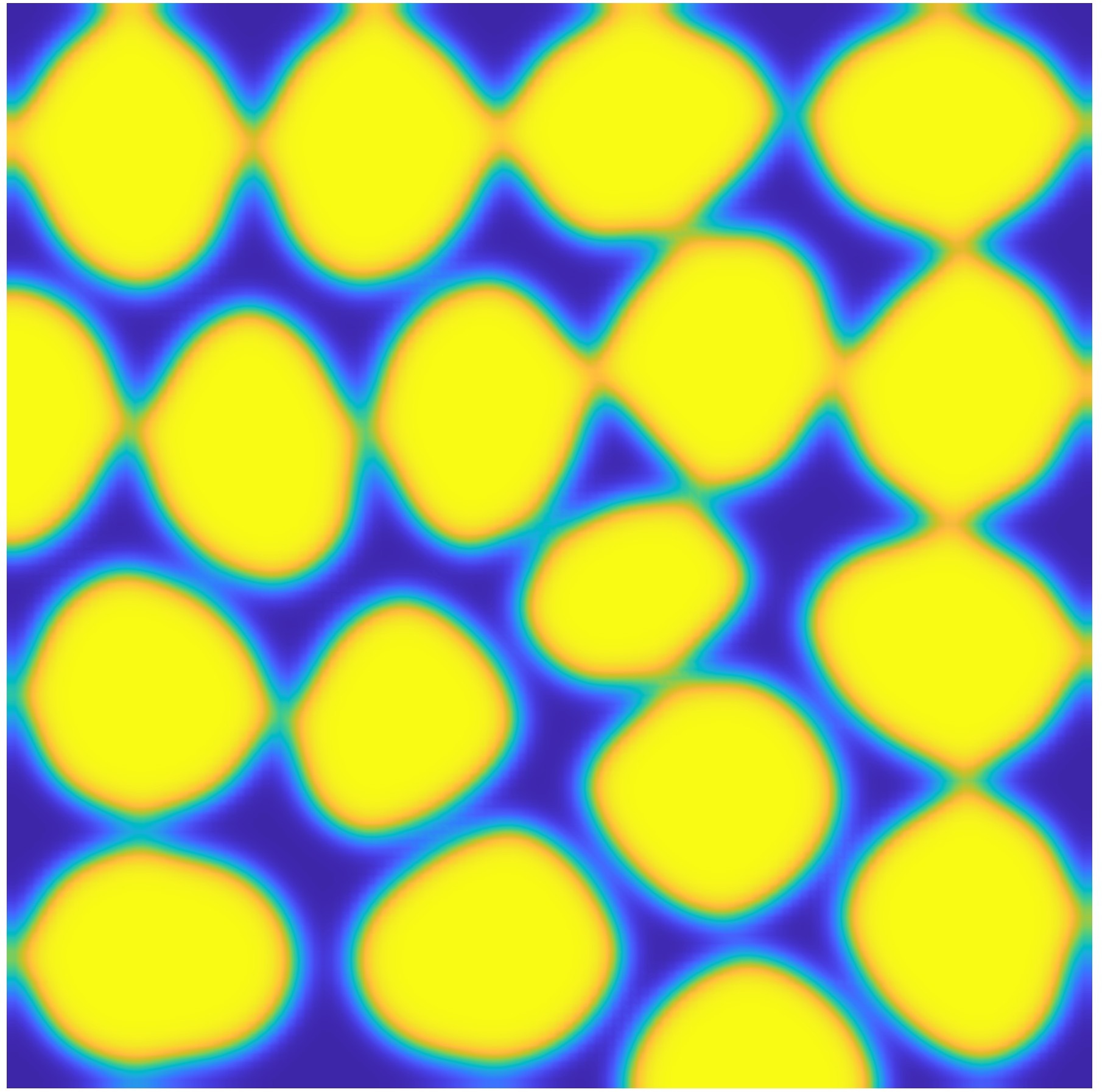}}\hfill
\subfloat[Angular]{\includegraphics[width=0.15\textwidth]{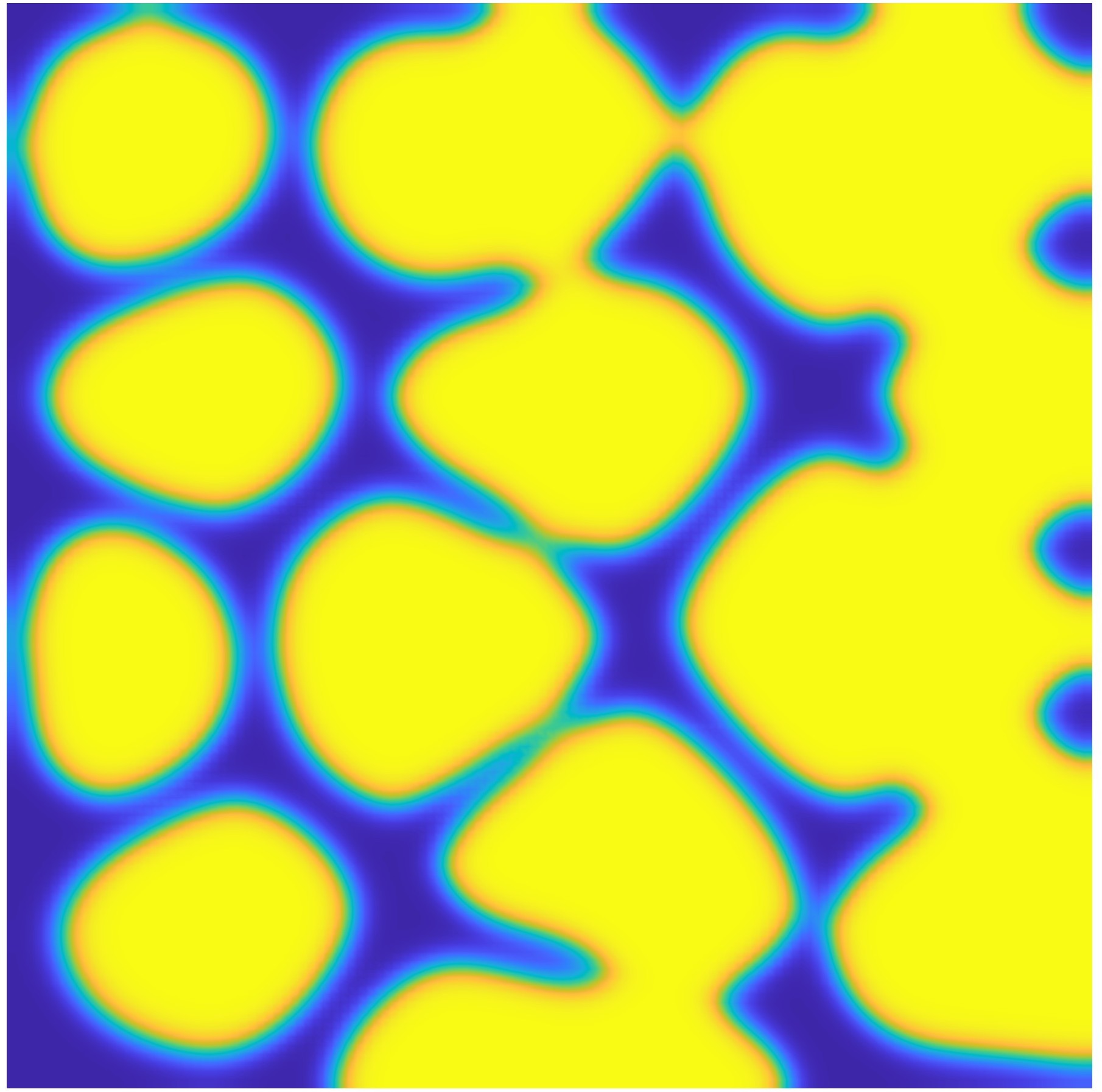}}\hfill
\subfloat[Curved]{\includegraphics[width=0.15\textwidth]{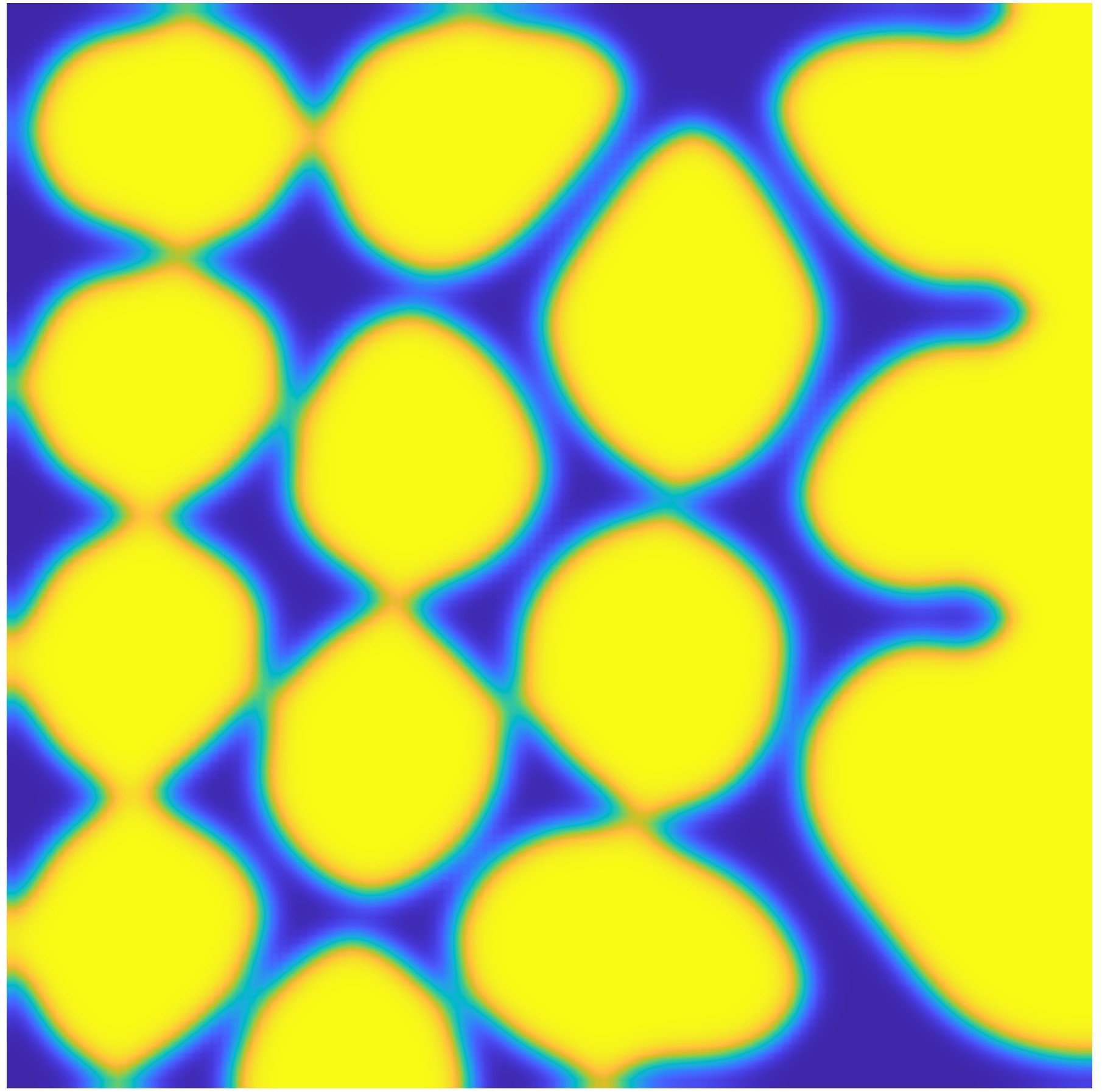}}\hfill
\subfloat[Pulsed]{\includegraphics[width=0.15\textwidth]{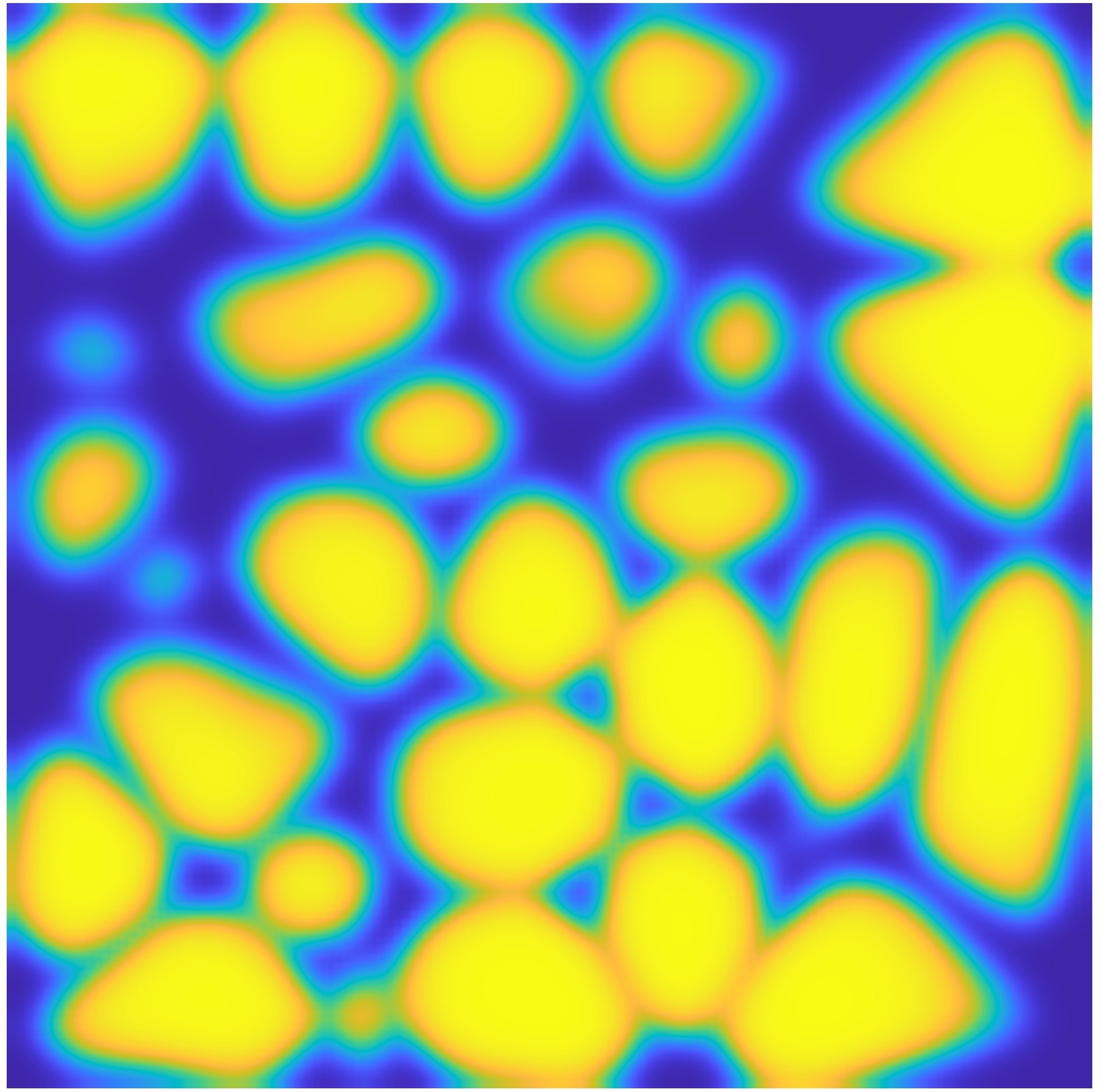}}
\caption{(Color online) We present the two-phase mushy zone structure in the $x$-$y$ plane corresponding to the bridging plane with zero Euler characteristic (\textit{i.e.}, $\chi$ = 0 in Eq.~\eqref{eq_euler}). The bridging planes demonstrate the effects of varying isotherm patterns on domain morphology. The solid is yellow, and the liquid is blue.}\label{fig_plane}
\end{figure}

\begin{figure}[ht]
\centering
\includegraphics[width=0.8\textwidth]{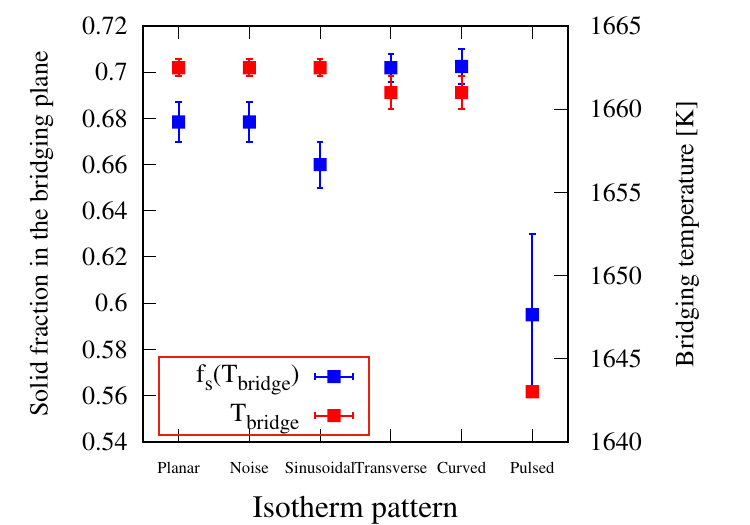}
\caption{We estimate the solid fraction ($f_s$) and bridging temperature ($T_\text{bridge}$) from the bridging plane depicted in Fig.~\ref{fig_plane} and plotted for various isotherms.}\label{fig_bridging}
\end{figure}

For a reasonably complete characterization of the bridging phenomena, we present the mushy zone structure in the bridging plane for all isotherm patterns (Fig.~\ref{fig_plane}). The solid and liquid phases in the mushy zone can be clearly identified, with solid bridges forming between the neighboring cells. Compared to the planar isotherm (Fig.~\ref{fig_plane}a), mushy zone evolution does not change when Gaussian noise disturbs the isotherm (Fig.~\ref{fig_plane}b). However, the domain morphology is significantly different for other isotherm patterns. For the sinusoidal isotherm, the bridging plane shows modified solidification paths with a different number of cells and their spatial arrangements (Fig.~\ref{fig_plane}c). A gradient of cell sizes forms with the transverse isotherm (Fig.~\ref{fig_plane}d). With the curved isotherm, the cell sizes vary depending on the magnitude of the curvature, with the strong curvature of the isotherm leading to massive cells (right in Fig.~\ref{fig_plane}e) within an otherwise nearly uniform domain morphology. For a comparative analysis, we plot the characteristic values of $f_s$ and $T$ in the bridging plane in Fig.~\ref{fig_bridging}, demonstrating the sensitivity of the temperature isotherms on the mushy zone structure and solidification path.

\begin{figure}[ht]
\centering
\subfloat[]{\includegraphics[scale=0.52]{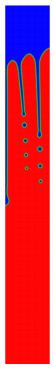}}\hspace{10mm}
\subfloat[]{\includegraphics[scale=0.3]{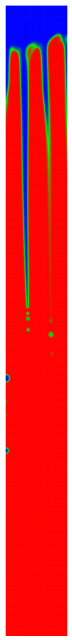}}\hspace{10mm}
\subfloat[]{\includegraphics[scale=0.8]{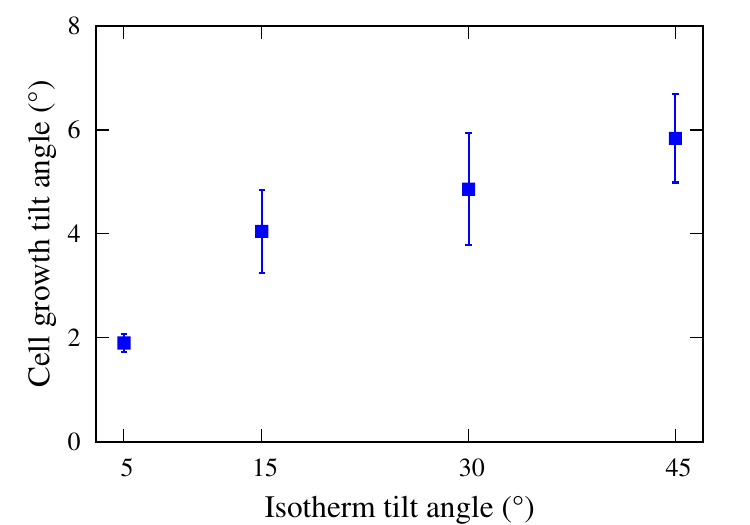}}
\caption{(Color online) Transverse isotherm leads to a tilted cellular growth pattern, as shown using (a) 2D simulations and (b) a 2D slice from 3D simulations. The simulations are run with identical parameters. The tilt angle is approximately the same ($\approx$4$^\circ$) in both simulations. In this phase-field ($\phi$) representation, the solid cell is red, and the liquid is blue. (c) From 2D simulations, cell growth angle is plotted as a function of the inclination angle of the transverse isotherm.}\label{fig_tilt}
\end{figure}

It is noteworthy that cellular structures tilt due to an inclination of the temperature isotherm (\textit{e.g.}, transverse) with respect to the planar setup (Fig.~\ref{fig_tilt}). Specifically, a competition between the heat flow direction established by the applied thermal gradient direction (\textit{i.e.}, process anisotropy) and the preferred crystalline direction (\textit{i.e.}, vertical) resulting from interfacial anisotropy ($a(\hat{n})$ in Eq.~\eqref{eq_anisotropy}) sets in, governing the cellular growth direction. Such tilted growth behavior has been commonly observed in experiments. While previous studies work with the anisotropy of the free energy of the solid-liquid interface, we show here, for the first time, that an inclination of the local isotherm can also generate titled arrays of cellular patterns under directional solidification conditions. An preliminary analysis of the growth tilt angle as a function of the inclination angle of the isotherm is shown in Fig.~\ref{fig_tilt}c. The tilt angle is much smaller in these cases compared to interfacial anisotropy-mediated tilted growth~\cite{tourret_2015,xing2015phase}.

\begin{figure}[htbp]
\centering
\includegraphics[width=\textwidth]{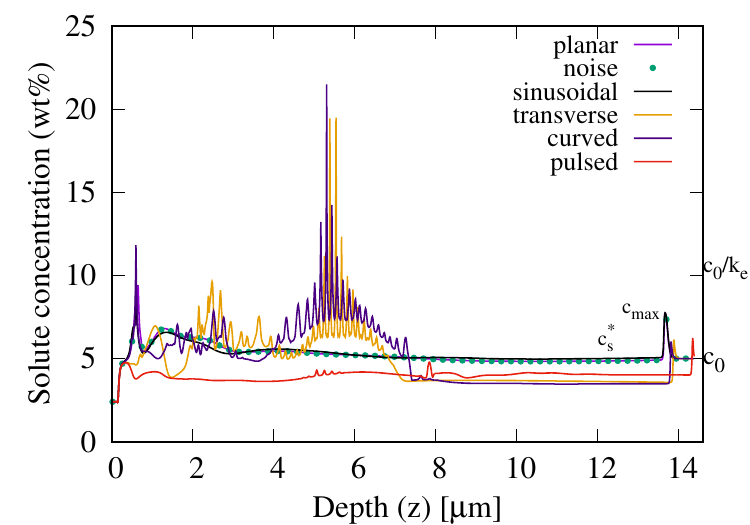}
\caption{(Color online) We show the time history of solute concentration profiles in the growth direction through a cell tip for various isotherm patterns. Significant differences in the concentration profiles arise from the effects of different isotherm patterns. Note that the curves with the planar isotherm with and without the noise show no difference. For the planar case, we mark the solid concentration ($c_s^*$) just behind the interfacial region and the maximum concentration ($c_{\text{max}}$) to calculate microsegregation ($k_v = c_s^*/c_{\text{max}}$) (Fig.~\ref{fig_segregation}). Similarly, $k_v$ can be calculated for other isotherms. The far-field liquid concentration is given by $c_0$ =  5 wt\% Nb. The oscillations in the concentration profiles in the solid correspond to the ``droplets'' containing solute-rich liquid. The fraction of droplets with a liquid concentration above $c_0/k_e$ is calculated to estimate the formation of segregation-induced secondary phases (Fig.~\ref{fig_laves}).}\label{fig_composition}
\end{figure}

\begin{figure}[htbp]
\centering
\includegraphics[width=0.75\textwidth]{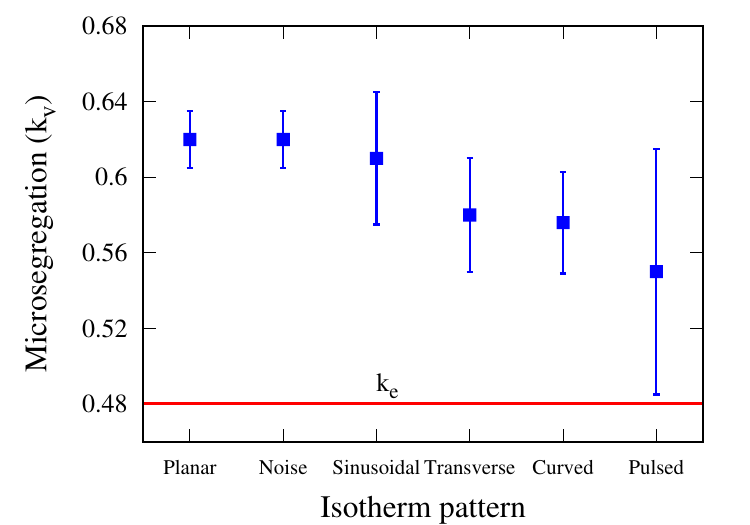}
\caption{Microsegregation ($k_v$) is calculated by the ratio of the solid concentration just behind the interfacial region to the maximum concentration from the concentration profiles generated with different isotherm patterns (see Fig.~\ref{fig_composition}). The equilibrium partition coefficient $k_e$ is shown for reference to show the deviation of solute profiles from local equilibrium. On average, non-planar isotherms produce a smaller $k_v$ compared to planar isotherms. Each $k_v$ is plotted with a confidence interval, representing the standard deviation around the mean obtained by averaging the data.}\label{fig_segregation}
\end{figure}

Next, we analyze the concentration profiles from the $x$-$y$ vertical sections. Visualization of the time history of concentration profiles is difficult in 3D. Therefore, for a meaningful representation of the solute profiles, we extract the traces of concentration from the microstructure evolutionary path along several lines in the growth direction $z$. The positions of these lines are chosen to pass directly through cell tips. Note that the solute diffusivity in the solid is taken as zero, allowing us to obtain the best definition of the solute profiles in the solid and liquid at the interface. We show the plots of concentration versus distance along a line in the growth direction for various isotherm patterns in Fig.~\ref{fig_composition}. The build-up of solute (``spike'' on the right) in the liquid ahead of the interface, as well as the profile produced in the solid, can be clearly seen. The concentration plateau in the solid just behind the interfacial region is taken as the cell core concentration ($c_s^*$). The spike in composition ($c_{\text{max}}$) in the liquid at the interface signifies the rejection of solute by the growing cells that enriches the remaining liquid between cells. Beyond $c_{\text{max}}$, the far-field liquid concentration $c_0$ (5 wt\% Nb) is obtained. The concentration patterns represented by the spikes in the solid (on the left) represent the trail of solute-rich liquid droplets that are shed from the root region of the cells. Analyzing the profiles of $c_s^*$ and $c_{\text{max}}$, it is evident that isotherm patterns influence the degree of solute partitioning across the interfacial region. We estimate the microsegregation by the ratio~\cite{dantzigbook,Aziz1982}, 
\begin{equation}
 k_v = \frac{c_s^*}{c_{\text{max}}},
\end{equation}
which varies with isotherm patterns (Fig.~\ref{fig_segregation}). The non-planar isotherms, particularly the pulsed isotherm, result in a smaller $k_v$ compared to planar isotherms (within the measurement uncertainty).

\begin{figure}[htbp]
\centering
\includegraphics[width=0.75\textwidth]{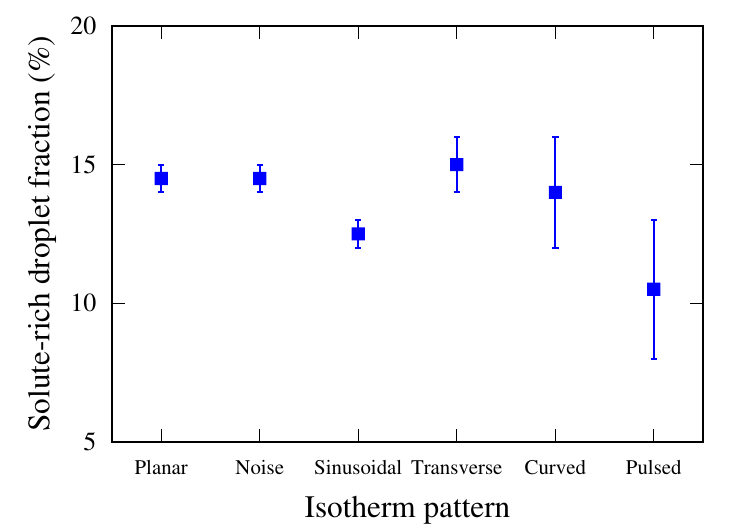}
\caption{The volume fraction of droplets containing the solute-rich liquid is estimated by analyzing the concentration profiles above $c_0/k_e$ (see Fig.~\ref{fig_composition}). The droplets could transform into secondary phases during the late stages of solidification.}\label{fig_laves}
\end{figure}

 It is straightforward to conjecture that the solute-rich liquid droplets will eventually freeze at a much lower temperature and could influence the formation of secondary phases, as is often seen experimentally (\textit{e.g.}, Liquid $\rightarrow$ $\gamma$ + Laves when approximating Ni-Nb as a quasi-binary Inconel 718 alloy)~\cite{nie_2014,wang2019investigation,Ghosh2018_droplets}. We use a heuristic approach to estimate how solute-rich areas transition to secondary phases by analyzing concentration profiles in the sample that exceed $c_0/k_e$ (beyond the equilibrium liquid concentration at the interface). The analysis is shown in Fig.~\ref{fig_laves}. Solute enrichment tends to be lower with sinusoidal and pulsed isotherms compared to other types of isotherms. These observations also agree with the reduced oscillations in the concentration patterns resulted in the solid (Fig.~\ref{fig_composition}). In addition, the number of droplets significantly decreases with the pulsed isotherm (laser-off period), which controls the transition from deep cells to shallow cells, reducing the overall solute enrichment in the mushy zone. This scenario could lead to a low tendency for defect formation and unexpected precipitation of secondary phases in the solidified material~\cite{tucho_2017,kuo2017effect,kouraytem2021,sun2022hottearing}. 

\section{Discussion}\label{sec:discussion}
We have shown by three-dimensional phase-field simulations that a disturbance in the thermal gradient profile, leading to various isotherm patterns, significantly affects the resulting cellular growth patterns. To our knowledge, there have been very few experiments and no comprehensive numerical studies on the role of isotherm patterns on microstructure evolution during melt pool solidification. Therefore, we could not make a quantitative comparison of our results with the literature. However, the generic cellular growth behavior obtained is consistent with the existing literature. A very fine columnar microstructure of $\gamma$- or $\alpha$ phase cells with $\lambda_c$ $\approx$ \SI{1}{\micro\meter} or below is usually observed in AM of stainless steels~\cite{haghdadi2021additive_review,rannar2017hierarchical}. The experiments with periodic perturbations produced by sinusoidal hatching resulted in overall refined cellular structures during the laser powder bed fusion of 316L stainless steels~\cite{sinusoidal_mussatto2022laser}. The experiments with pulsed laser-induced perturbations led to fine-sized $\gamma$-cells, reduced microsegregation, enhanced possibility of the columnar-to-equiaxed transition (CET), the reduction of secondary phases, and the formation of tilted dendrites in Ni-Nb alloys~\cite{li2017melt_pulsed}. The pulsed laser refined the prior-$\beta$ columnar structures and promoted CET during the AM of Ti-6Al-4V~\cite{pulsed_cai2024grain,pulsed_yoon2022}. 

Our results are in good agreement with the above experiments, taking into account that the simulated microstructure features are expected to be slightly different from the actual situations during fabrication, due to the different alloy systems and control parameters between simulations and AM experiments. Our simulations show extremely fine cellular structures with $\lambda_c$ < \SI{1}{\micro\meter}, as commonly observed in the as-solidified AM specimens. On average, $\lambda_c$ is minimum when using non-planar isotherms (particularly the sinusoidal and pulsed) with $\lambda_c$ $\approx$ \SI{0.06}{\micro\meter}. The analysis of $f_s$ in the vertical sections indicates that the mushy zone is smaller for the planar case, while non-planar isotherms (particularly the transverse and curved) produce a longer mushy zone. The longer the mushy zone, the last residual liquid (\textit{i.e.}, droplets) in the mushy zone is smaller in amount and is even further enriched with the solute, increasing the tendency for forming segregation-induced defects during terminal solidification. The concentration profiles indicate that the extent of solute partitioning across the interface (\textit{i.e.}, $k_v$) decreases for non-planar isotherms (particularly the sinusoidal and pulsed). Moreover, some level of solute trapping should be physically present due to the rapid AM solidification, as observed in our simulations, consistent with experimental observations~\cite{karayagiz2020,Trevor2017}. Overall, our results show that the isotherm patterns significantly affect $\lambda_c$, $f_s$, and $k_v$. However, we have found no general relationship between the isotherm patterns and the resulting cellular features. While these observations are crucial in understanding the macroscopic behavior of the mushy zones, they have been little studied so far because they are difficult to visualize and hence characterize from experiments. When considering the effects of varying isotherm parameters, it is straightforward to understand that as the geometry of the isotherm patterns approaches the planar pattern, the differences between their effects will gradually reduce, and an identical growth behavior will eventually result. Moreover, multigrain simulations in which a growth competition between multiple grains under various isotherm patterns will contribute to more physical insights on cellular growth in real materials; however, such 3D simulations require larger domains and are not possible with our computational resources. 

Our results reasonably agree with recent experiments and simulations with pulsed beam profiles that led to cellular morphology with smaller cell sizes and lesser microsegregation and, hence, reduced secondary phase formation~\cite{pulsed_cai2024grain,pulsed_yoon2022,li2017melt_pulsed,yang2024phase}. In experiments, the segregation-induced secondary phases could minimize the occurrence of solidification defects by grain refinement~\cite{durga2021grain,julien2021nucleation,sun2022hottearing}. In view of these observations, our simulations could provide more clues about designing the heat treatment schedules to control the microsegregation and solute trapping behavior observed during non-equilibrium solidification of melt pools. In simulations with pulsed isotherm, deep cellular shapes that develop during the laser-on period transform into small-amplitude shallow cells during the laser-off period. Therefore, we do not observe a multi-directional solidification phenomenon (\textit{i.e.}, equiaxed growth), as reported in Refs.~\cite{li2017melt_pulsed,samy2023_cet}, likely due to the absence of nucleating particles in the melt. In this context, the CET would make an interesting future work, where heterogeneous nuclei added in the melt ahead of the cell tips could transform into elongated equiaxed dendrites under directional solidification~\cite{durga2021grain,samy2023_cet,julien2021nucleation,nabavizadeh2020_cet}. Our preliminary simulations suggest that an inclination of the isotherm pattern produces tilted growth of the cellular patterns. Naturally, the free energy anisotropy of the solid-liquid interface (\textit{i.e.}, interfacial anisotropy in Eq.~\eqref{eq_anisotropy}) controls the crystallographic orientations in a columnar structure. Thus, an orientation selection competition between different isotherm patterns and interfacial anisotropy would be interesting to explore to simulate possible growth directions and morphologies of the cellular interface. Work in this particular direction is currently in progress.

Compared to 2D phase-field simulations, 3D simulations generally come with the additional cost of significantly high computing time and data storage requirements and considerable complications in the interpretability of the obtained results. The additional dimension (\textit{i.e.}, space degrees of freedom) makes the solute diffusion very efficient in 3D, producing more realistic ``numerical experiments'' and better agreement with experiments. However, the problem of melt convection arises in 3D. Notably, 3D simulations of dilute binary alloy solidification (similar to our study) demonstrated the negligible influence of convection on solute partitioning across the interfacial region~\cite{Lee2010}. The effects of convection on cellular shapes are not as pronounced as those on dendrites (\textit{i.e.}, cells with lateral branches), which we do not observe in our simulations. Moreover, very dense cellular sub-structures (\textit{e.g.}, $\lambda_c$ < \SI{0.2} {\micro\meter} as in our study) could lead to significant resistance to fluid flow in experiments with an exponential increase of the damping effect in mushy zones~\cite{dantzigbook,damping_mushy}. Ignoring convection as a first-order approximation is therefore reasonable in this study. We note that compared to subgrain structures~\cite{yu2022impact,tang2022phase}, profound effects of melt pool convection have been reported on grain-scale microstructure evolution~\cite{xiong2022evaluate,rai2016ca}, which we do not simulate here. However, these effects should be considered for more accurate simulations by coupling the phase-field method with a Lattice Boltzmann~\cite{rai2016ca} or computational fluid dynamics~\cite{xiong2022evaluate} approach.

Finally, we must admit that we work in the solidification regime representative of AM, where complex physical interactions involving melt convection~\cite{wu2024unified}, interface kinetics~\cite{kinetic_effects}, and coupled thermal-solutal effects~\cite{li2023modeling} may be present, which we ignore as a first approximation. Clearly, a few extensions along these lines in the phase-field model components, including a treatment of the antitrapping current~\cite{mullis2010antitrapping} and curvature effects~\cite{kinetic_effects}, are necessary to get a more quantitative picture. In addition, a multicomponent, multiphase model formulation~\cite{radhakrishnan2018_3d,nestler2011_opinion} will allow simulations of the segregation-induced secondary phases from the ``droplets'' in the mushy zone at the end of solidification. Also, we assume that the net effects of either $G$ or $V$ on cellular growth through the cooling rate ($GV$) are the same; hence, we do not vary the growth velocity pattern in our simulations, but it may be essential to consider in AM solidification~\cite{babu_velocity_pattern}. Moreover, the effects of varying $G$ and $V$ in addition to varying their patterns will make a more complete future study; however, the large number of parameters and the complexity of the non-uniform distribution of solid and liquid phases and their concentration profiles will make the procedure more cumbersome and interpretation of the results more challenging. Since we keep all the numerical, material, and cooling parameters fixed with only variation in the temperature isotherms, the relative effects of the aforementioned phase-field model enhancements on the influence of isotherm patterns could be considered the same. Therefore, simulations have been performed, at least qualitatively, with reasonable approximations for the average cellular growth behavior under various isotherm patterns far above the constitutional supercooling threshold representative of the melt pool solidification. 

\section{Conclusions and outlook}\label{sec:summary}
From this work, we conclude:

\begin{itemize}
\item The cellular growth patterns can be significantly influenced by the isotherm patterns generated in the solidifying melt pools. The planar and non-planar isotherms produce considerably different solidification and phase segregation features.

\item The solid fraction ($f_s$) varies as a function of depth ($z$) with isotherm patterns (Fig.~\ref{fig_fraction}), with maximum difference of $\approx$ 60\% has been observed between the steepness of $f_s$ curves (\textit{i.e.}, $df_s/dz$), signifying different extensions of the mushy zone. Non-planar isotherms typically produce a smaller $df_s/dz$ and, hence, a longer mushy zone compared to the planar case.

\item Solid percolation results in the mushy zone for $f_s$ between 0.6 and 0.8 for various isotherm patterns (Fig.~\ref{fig_bridging}), with a maximum difference of $\approx$ 20\% observed between planar and other isotherms, suggesting that the transition from a liquid-like to a solid-like behavior of the material takes place at different mushy zone planes for different isotherm patterns. 

\item Different isotherm patterns generate compositional variations of varying degrees across the interfacial region, with a maximum difference of $\approx$ 14\% has been estimated in microsegregation ($k_v$) between planar and other isotherms, indicating different segregation patterns in the solidified material. Non-planar isotherms generally result in a smaller $k_v$ than the planar isotherm (Fig.~\ref{fig_segregation}).

\item The cell spacing ($\lambda_c$) can be roughly approximated from the Euler characteristic ($\chi$) curve (Fig.~\ref{fig_connection}), with its peak indicates the number of cells in the simulation box. Thus, different isotherm patterns result in various cell sizes, with a maximum difference of $\approx$ 15\% has been observed between planar and other isotherms. On average, the sinusoidal and pulsed isotherms produce a finer columnar structure.

\item The volume fraction of segregation-induced liquid droplets that are detached from the bottom of the grooves between the cells varies with isotherm patterns (Fig.~\ref{fig_laves}), with a maximum difference of $\approx$ 40\% results between the solute profiles in the droplets generated by different isotherms, indicating their potential influence on the type and amount of secondary phases that could develop during subsequent evolution.
\end{itemize}

The above points indicate that variations in the isotherm pattern could lead to different microstructure descriptors (\textit{e.g.}, $\phi$, $c$, $T$) and features (\textit{e.g.}, $\lambda_c$, $f_s$, $\chi$, $k_v$). Thus, for a more quantitative analysis, this variation should be considered for modeling AM solidification or as a potential source of uncertainty to the statistical variation in microstructural features. Since the phase-field simulations of cellular structures on a full-melt pool scale in 3D with very thin interfaces (\textit{i.e.}, small $W_0$) are computationally prohibitive, our approach provides a reasonable alternative for performing simulations with different isotherm patterns corresponding to the geometry of different melt pool boundary segments and subsequently interpolating the results to obtain a reasonably complete picture of melt pool solidification. We must admit that our approach uses strongly idealized approximations to model rapid solidification, realizing a baseline assessment of cellular growth under AM conditions. Nevertheless, the present study demonstrates a more general theoretical approach for the first time to provide clues about fabrication trials with appropriate process adjustment by heat input patterns or other non-equilibrium transient effects resulting in a melt pool disturbance for local microstructure control in AM and directional solidification in general. 

%----------------------------------------------------------------------------------------
\section*{CRediT authorship contribution statement}
\textbf{Saurabh Tiwari}: Methodology, Software, Validation, Formal Analysis, Data Curation. \textbf{Supriyo Ghosh}: Supervision, Project Administration, Funding Acquisition, Writing - Original Draft, Review \& Editing. 

\section*{Declaration of competing interest}
The authors have no conflicts to disclose.

\section*{Acknowledgments}
The supports from FIG (SRIC office, IIT Roorkee) and SERB (Government of India) are appreciated.

\section*{Data Availability Statement}
The research data are available from the corresponding author upon reasonable request.

\appendix

\section{Typical effects of temporal variation of $G$ induced by pulsed beam}\label{sec_appendix1}
\begin{figure}[htbp]
\centering
\subfloat[]{\includegraphics[width=0.48\textwidth]{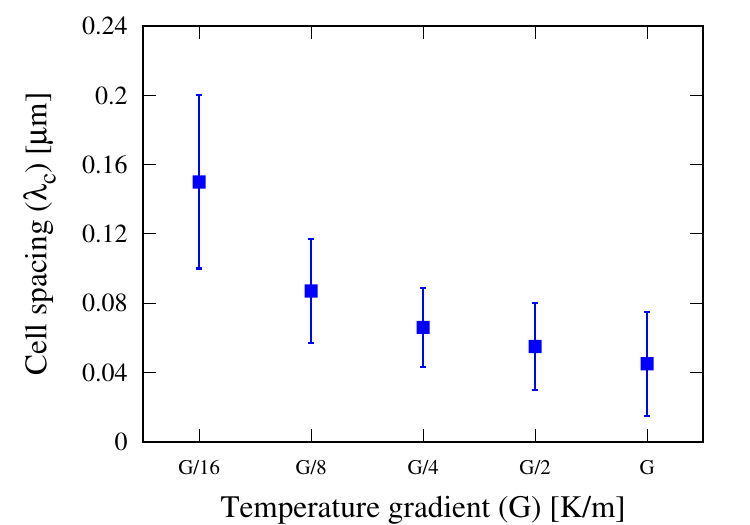}}
\subfloat[]{\includegraphics[width=0.48\textwidth]{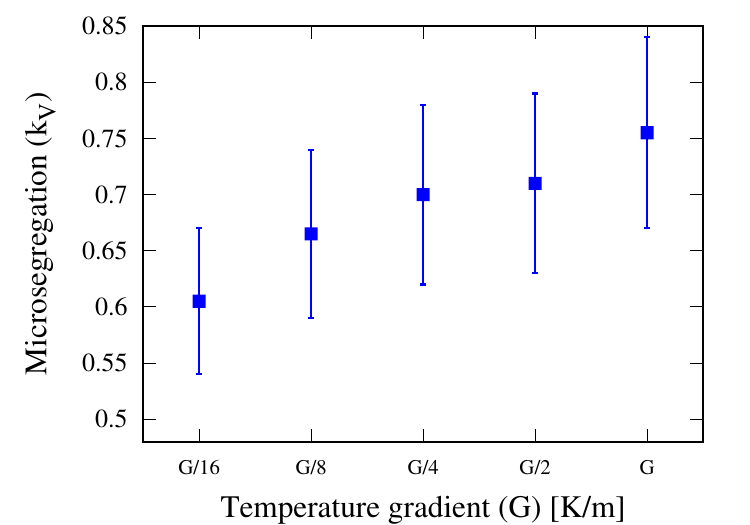}}
\caption{Simulation of the situation when the temperature gradient becomes a decreasing function of time; that is, $G$ periodically drops to a certain value after a given time interval (10000 $\Delta t$) during the laser-off period. Initial value of $G$ = $10^7$ K/m. Typical variations in (a) cell spacing and (b) microsegregation are shown. We note that the growth front does not reach a steady state for the small simulation time allowed for cells to grow for each $G$. Different values of $G$ are implemented by the planar isotherm following the pulse pattern.}\label{fig_gradient_time}
\end{figure}

Let us assume that instead of $G$ dropping to zero during the laser-off period, $G$ uniformly decreases over time (or increases during the laser-on period). The effects of such pulse $G$ patterns on cellular features are shown in Fig.~\ref{fig_gradient_time}. Depending on the residence time for each value of $G$, the simulations may not reach a steady state; however, the general trends are expected to hold within the long-time limit. These preliminary observations suggest that varying $G$ patterns resulting from pulse parameters can lead to considerable control over cellular solidification and phase segregation. These results could be highly desirable, allowing essentially independent control of the scale and extensions of the cellular structures and mushy zones, thereby enabling the site-specific microstructure control according to the local performance requirements of the component. For example, a structure may require fatigue resistance at one location given by fine cells (small $\lambda_c$) and creep resistance at another by coarse cells (large $\lambda_c$), as demonstrated in~\cite{moat2009crystallographic}. Indeed, various pulse patterns can be simulated for a very long time using larger domain sizes; however, the results could not be obtained within a reasonable time with our computational resources. Finally, using various pulsed $G$ patterns, where each value of $G$ contributes to site-specific control of the scale and extensions of the solidification features, a hierarchical microstructure will eventually develop in the final solid material.



\begin{thebibliography}{}
\expandafter\ifx\csname url\endcsname\relax
  \def\url#1{\texttt{#1}}\fi
\expandafter\ifx\csname urlprefix\endcsname\relax\def\urlprefix{URL }\fi
\expandafter\ifx\csname href\endcsname\relax
  \def\href#1#2{#2} \def\path#1{#1}\fi

\end{thebibliography}


\section*{References}
\begin{thebibliography}{10}
\expandafter\ifx\csname url\endcsname\relax
  \def\url#1{\texttt{#1}}\fi
\expandafter\ifx\csname urlprefix\endcsname\relax\def\urlprefix{URL }\fi
\expandafter\ifx\csname href\endcsname\relax
  \def\href#1#2{#2} \def\path#1{#1}\fi

\bibitem{debroy_additive}
T.~DebRoy, H.~Wei, J.~Zuback, T.~Mukherjee, J.~Elmer, J.~Milewski, A.~Beese,
  A.~Wilson-Heid, A.~De, W.~Zhang, Additive manufacturing of metallic
  components--process, structure and properties, Progress in Materials Science
  92 (2018) 112--224.

\bibitem{liu2022additive}
Z.~Liu, D.~Zhao, P.~Wang, M.~Yan, C.~Yang, Z.~Chen, J.~Lu, Z.~Lu, Additive
  manufacturing of metals: Microstructure evolution and multistage control,
  Journal of Materials Science \& Technology 100 (2022) 224--236.

\bibitem{haghdadi2021additive_review}
N.~Haghdadi, M.~Laleh, M.~Moyle, S.~Primig, Additive manufacturing of steels: a
  review of achievements and challenges, Journal of Materials Science 56 (2021)
  64--107.

\bibitem{review_meltpool}
J.~Wang, R.~Zhu, Y.~Liu, L.~Zhang, Understanding melt pool characteristics in
  laser powder bed fusion: An overview of single-and multi-track melt pools for
  process optimization, Advanced Powder Materials 2~(4) (2023) 100137.

\bibitem{sanchez2021_review}
S.~Sanchez, P.~Smith, Z.~Xu, G.~Gaspard, C.~J. Hyde, W.~W. Wits, I.~A.
  Ashcroft, H.~Chen, A.~T. Clare, Powder bed fusion of nickel-based
  superalloys: A review, International Journal of Machine Tools and Manufacture
  165 (2021) 103729.

\bibitem{Trevor2017}
T.~Keller, G.~Lindwall, S.~Ghosh, L.~Ma, B.~Lane, F.~Zhang, U.~R. Kattner,
  E.~A. Lass, J.~C. Heigel, Y.~Idell, M.~E. Williams, A.~J. Allen, J.~E. Guyer,
  L.~E. Levine, Application of {F}inite {E}lement, {P}hase-field, and
  {CALPHAD}-based {M}ethods to {A}dditive {M}anufacturing of {N}i-based
  {S}uperalloys, Acta {M}aterialia 139 (2017) 244--253.

\bibitem{karayagiz2020}
K.~Karayagiz, L.~Johnson, R.~Seede, V.~Attari, B.~Zhang, X.~Huang, S.~Ghosh,
  T.~Duong, I.~Karaman, A.~Elwany, et~al., Finite interface dissipation phase
  field modeling of {Ni-Nb} under additive manufacturing conditions, Acta
  Materialia 185 (2020) 320--339.

\bibitem{hecht2019am}
G.~Boussinot, M.~Apel, J.~Zielinski, U.~Hecht, J.~Schleifenbaum, Strongly
  out-of-equilibrium columnar solidification during laser powder-bed fusion in
  additive manufacturing, Physical Review Applied 11~(1) (2019) 014025.

\bibitem{acharya2017prediction}
R.~Acharya, J.~A. Sharon, A.~Staroselsky, Prediction of microstructure in laser
  powder bed fusion process, Acta Materialia 124 (2017) 360--371.

\bibitem{Mullins1964}
W.~W. Mullins, R.~F. Sekerka, Stability of a planar interface during
  solidification of a dilute binary alloy, Journal of Applied Physics 35~(2)
  (1964) 444--451.

\bibitem{dantzigbook}
M.~Rappaz, J.~A. Dantzig, Solidification, EFPL Press, Lausanne, 2009.

\bibitem{li2017melt_pulsed}
S.~Li, H.~Xiao, K.~Liu, W.~Xiao, Y.~Li, X.~Han, J.~Mazumder, L.~Song, Melt-pool
  motion, temperature variation and dendritic morphology of inconel 718 during
  pulsed-and continuous-wave laser additive manufacturing: A comparative study,
  Materials \& design 119 (2017) 351--360.

\bibitem{scan_sun2018effect}
S.-H. Sun, K.~Hagihara, T.~Nakano, Effect of scanning strategy on texture
  formation in {Ni-25 at.\% Mo} alloys fabricated by selective laser melting,
  Materials \& Design 140 (2018) 307--316.

\bibitem{yao2023melt}
L.~Yao, Z.~Xiao, A.~Ramesh, Y.~Zhang, On the melt pool flow and interface shape
  of dissimilar alloys via selective laser melting, International
  Communications in Heat and Mass Transfer 145 (2023) 106833.

\bibitem{mathis_transverse}
M.~Perrut, A.~Parisi, S.~Akamatsu, S.~Bottin-Rousseau, G.~Faivre, M.~Plapp,
  Role of transverse temperature gradients in the generation of lamellar
  eutectic solidification patterns, Acta Materialia 58~(5) (2010) 1761--1769.

\bibitem{dinda2013}
G.~Dinda, A.~Dasgupta, S.~Bhattacharya, H.~Natu, B.~Dutta, J.~Mazumder,
  Microstructural characterization of laser-deposited al 4047 alloy,
  Metallurgical and Materials Transactions A 44~(5) (2013) 2233--2242.

\bibitem{debroy2015texture}
H.~Wei, J.~Mazumder, T.~DebRoy, Evolution of solidification texture during
  additive manufacturing, Scientific reports 5~(1) (2015) 1--7.

\bibitem{scan_khairallah2020ca}
R.~Shi, S.~A. Khairallah, T.~T. Roehling, T.~W. Heo, J.~T. McKeown, M.~J.
  Matthews, Microstructural control in metal laser powder bed fusion additive
  manufacturing using laser beam shaping strategy, Acta Materialia 184 (2020)
  284--305.

\bibitem{scan_roehling2020controlling}
T.~T. Roehling, R.~Shi, S.~A. Khairallah, J.~D. Roehling, G.~M. Guss, J.~T.
  McKeown, M.~J. Matthews, Controlling grain nucleation and morphology by laser
  beam shaping in metal additive manufacturing, Materials \& Design 195 (2020)
  109071.

\bibitem{ebrahimi2023revealing}
A.~Ebrahimi, M.~Sattari, A.~Babu, A.~Sood, G.-W.~R. R{\"o}mer, M.~J. Hermans,
  Revealing the effects of laser beam shaping on melt pool behaviour in
  conduction-mode laser melting, Journal of Materials Research and Technology
  27 (2023) 3955--3967.

\bibitem{chen2024situ}
Y.~Chen, D.~Zhang, P.~O’Toole, D.~Qiu, M.~Seibold, K.~Schricker, J.-P.
  Bergmann, A.~Rack, M.~Easton, In situ observation and reduction of hot-cracks
  in laser additive manufacturing, Communications Materials 5~(1) (2024) 84.

\bibitem{sinusoidal_mussatto2022laser}
A.~Mussatto, R.~Groarke, R.~K. Vijayaraghavan, M.~A. Obeidi, P.~J. McNally,
  V.~Nicolosi, Y.~Delaure, D.~Brabazon, Laser-powder bed fusion of silicon
  carbide reinforced 316l stainless steel using a sinusoidal laser scanning
  strategy, Journal of Materials Research and Technology 18 (2022) 2672--2698.

\bibitem{pulsed_cai2024grain}
Y.~Cai, Z.~Peng, J.~Chen, H.~Chen, J.~Xiong, Grain refinement and anisotropy
  improvement of arc-directed energy deposited {Ti-6Al-4V} with oscillating
  laser, Materials Science and Engineering: A 893 (2024) 146144.

\bibitem{ghosh2017primary}
S.~Ghosh, L.~Ma, N.~Ofori-Opoku, J.~E. Guyer, On the primary spacing and
  microsegregation of cellular dendrites in laser deposited {Ni-Nb} alloys,
  Model. Simul. Mater. Sci. Eng. 25~(6) (2017) 065002.

\bibitem{ghosh2022tusas}
S.~Ghosh, C.~K. Newman, M.~M. Francois, Tusas: A fully implicit parallel
  approach for coupled phase-field equations, Journal of Computational Physics
  448 (2022) 110734.

\bibitem{ghosh2018single}
S.~Ghosh, L.~Ma, L.~E. Levine, R.~E. Ricker, M.~R. Stoudt, J.~C. Heigel, J.~E.
  Guyer, Single-track melt-pool measurements and microstructures in {Inconel}
  625, {JOM} 70~(6) (2018) 1011--1016.

\bibitem{ghosh2019uncertainty}
S.~Ghosh, M.~Mahmoudi, L.~Johnson, A.~Elwany, R.~Arroyave, D.~Allaire,
  Uncertainty analysis of microsegregation during laser powder bed fusion,
  Model. Simul. Mater. Sci. Eng. 27~(3) (2019) 034002.

\bibitem{ghosh2020statistical}
S.~Ghosh, R.~Seede, J.~James, I.~Karaman, A.~Elwany, D.~Allaire, R.~Arroyave,
  Statistical modelling of microsegregation in laser powder-bed fusion,
  Philosophical Magazine Letters 100~(6) (2020) 271--282.

\bibitem{kang2024highly}
S.~B. Kang, G.~Huang, G.~Singhal, D.~Xie, D.~H. Hsieh, Y.~Lee, A.~A. Kulkarni,
  J.~W. Smith, Q.~Chen, K.~Thornton, et~al., Highly ordered eutectic
  mesostructures via template-directed solidification within thermally
  engineered templates, Advanced Materials 36~(15) (2024) 2308720.

\bibitem{zhang2022_shock}
A.~Zhang, Z.~Guo, B.~Jiang, J.~Song, F.~Pan, S.~Xiong, Effect of laser shock on
  lamellar eutectic growth: {A} phase-field study, International Journal of
  Heat and Mass Transfer 183 (2022) 122069.

\bibitem{boettinger2002}
W.~J. Boettinger, J.~A. Warren, C.~Beckermann, A.~Karma, Phase-field simulation
  of solidification, Annu. Rev. Mater. Res. 32 (2002) 163--194.

\bibitem{steinbach2009}
I.~Steinbach, Phase-field models in materials science, Model. Simul. Mater.
  Sci. Eng. 17 (2009) 073001.

\bibitem{chen2024_review}
L.-Q. Chen, N.~Moelans, Phase-field method of materials microstructures and
  properties, MRS Bulletin 49 (2024) 551--555.

\bibitem{Echebarria2004}
B.~Echebarria, R.~Folch, A.~Karma, M.~Plapp, Quantitative phase-field model of
  alloy solidification, Physical {R}eview {E} 70~(6) (2004) 061604.

\bibitem{ghosh2018ti64}
S.~Ghosh, K.~McReynolds, J.~E. Guyer, D.~Banerjee, Simulation of temperature,
  stress and microstructure fields during laser deposition of {Ti-6Al-4V},
  Model. Simul. Mater. Sci. Eng. 26~(7) (2018) 075005.

\bibitem{ghosh20183d}
S.~Ghosh, N.~Ofori-Opoku, J.~E. Guyer, Simulation and analysis of $\gamma$-{Ni}
  cellular growth during laser powder deposition of {Ni}-based superalloys,
  Computational Materials Science 144 (2018) 256--264.

\bibitem{chouhan2024modeling}
A.~Chouhan, L.~M{\"a}dler, N.~Ellendt, Modeling of rapid solidification in
  laser powder bed fusion processes, Computational Materials Science 238 (2024)
  112918.

\bibitem{sahoo2016phase}
S.~Sahoo, K.~Chou, Phase-field simulation of microstructure evolution of
  {Ti-6Al-4V} in electron beam additive manufacturing process, Additive
  manufacturing 9 (2016) 14--24.

\bibitem{farzadi2008}
A.~Farzadi, M.~Do-Quang, S.~Serajzadeh, A.~Kokabi, G.~Amberg, Phase-field
  simulation of weld solidification microstructure in an {Al-Cu} alloy, Model.
  Simul. Mater. Sci. Eng. 16~(6) (2008) 065005.

\bibitem{wang2019investigation}
X.~Wang, P.~Liu, Y.~Ji, Y.~Liu, M.~Horstemeyer, L.~Chen, Investigation on
  microsegregation of {IN718} alloy during additive manufacturing via
  integrated phase-field and finite-element modeling, Journal of Materials
  Engineering and Performance 28 (2019) 657--665.

\bibitem{karma2001}
A.~Karma, Phase-field formulation for quantitative modeling of alloy
  solidification, Physical Review Letters 87~(11) (2001) 115701.

\bibitem{knorovsky1989inconel}
G.~Knorovsky, M.~Cieslak, T.~Headley, A.~Romig, W.~Hammetter, Inconel 718: a
  solidification diagram, Metallurgical transactions A 20~(10) (1989)
  2149--2158.

\bibitem{provatasbook}
N.~Provatas, K.~Elder, Phase-field methods in materials science and
  engineering, John Wiley \& Sons, 2011.

\bibitem{ghosh2023_review}
S.~Ghosh, J.~Zollinger, M.~Zaloznik, D.~Banerjee, C.~K. Newman, R.~Arroyave,
  Modeling of hierarchical solidification microstructures in metal additive
  manufacturing: Challenges and opportunities, Additive Manufacturing 78 (2023)
  103845.

\bibitem{nie_2014}
P.~Nie, O.~Ojo, Z.~Li, Numerical modeling of microstructure evolution during
  laser additive manufacturing of a nickel-based superalloy, Acta Materialia 77
  (2014) 85--95.

\bibitem{asmdatabase}
T.~Massalski, H.~Okamoto, P.~Subramanian, L.~Kacprzak, Binary Alloy Phase
  Diagrams, ASM International, Materials Park, Ohio, 1990.

\bibitem{tucho_2017}
W.~M. Tucho, P.~Cuvillier, A.~Sjolyst-Kverneland, V.~Hansen, Microstructure and
  hardness studies of {Inconel} 718 manufactured by selective laser melting
  before and after solution heat treatment, Materials Science and Engineering:
  A 689 (2017) 220--232.

\bibitem{kuo2017effect}
Y.-L. Kuo, S.~Horikawa, K.~Kakehi, The effect of interdendritic $\delta$ phase
  on the mechanical properties of alloy 718 built up by additive manufacturing,
  Materials \& Design 116 (2017) 411--418.

\bibitem{kouraytem2021}
N.~Kouraytem, J.~Varga, B.~Amin-Ahmadi, H.~Mirmohammad, R.~A. Chanut, A.~D.
  Spear, O.~T. Kingstedt, A recrystallization heat-treatment to reduce
  deformation anisotropy of additively manufactured {Inconel} 718, Materials \&
  Design 198 (2021) 109228.

\bibitem{sun2022hottearing}
Z.~Sun, Y.~Ma, D.~Ponge, S.~Zaefferer, E.~A. J{\"a}gle, B.~Gault, A.~D.
  Rollett, D.~Raabe, Thermodynamics-guided alloy and process design for
  additive manufacturing, Nature Communications 13~(1) (2022) 1--12.

\bibitem{hoshen1976percolation}
J.~Hoshen, R.~Kopelman, Percolation and cluster distribution. {I.} cluster
  multiple labeling technique and critical concentration algorithm, Physical
  Review B 14~(8) (1976) 3438.

\bibitem{percolation_book}
D.~Stauffer, A.~Aharony, Introduction to percolation theory, Taylor \& Francis,
  London, UK, 2018.

\bibitem{spacing_3d}
S.~N. Tewari, Y.-H. Weng, G.~Ding, R.~Trivedi, Cellular array morphology during
  directional solidification, Metallurgical and Materials Transactions A 33
  (2002) 1229--1243.

\bibitem{kou2015criterion}
S.~Kou, A criterion for cracking during solidification, Acta Materialia 88
  (2015) 366--374.

\bibitem{tourret_2015}
D.~Tourret, A.~Karma, Growth competition of columnar dendritic grains: A
  phase-field study, Acta Materialia 82 (2015) 64--83.

\bibitem{xing2015phase}
H.~Xing, X.~Dong, C.~Chen, J.~Wang, L.~Du, K.~Jin, Phase-field simulation of
  tilted growth of dendritic arrays during directional solidification,
  International Journal of Heat and Mass Transfer 90 (2015) 911--921.

\bibitem{Aziz1982}
M.~J. Aziz, Model for solute redistribution during rapid solidification,
  Journal of Applied Physics 53~(2) (1982) 1158--1168.

\bibitem{Ghosh2018_droplets}
S.~Ghosh, M.~R. Stoudt, L.~E. Levine, J.~E. Guyer, Formation of {N}b-rich
  droplets in laser deposited {N}i-matrix microstructures, Scripta Materialia
  146 (2018) 36--40.

\bibitem{rannar2017hierarchical}
L.-E. R{\"a}nnar, A.~Koptyug, J.~Ols{\'e}n, K.~Saeidi, Z.~Shen, Hierarchical
  structures of stainless steel {316L} manufactured by electron beam melting,
  Additive Manufacturing 17 (2017) 106--112.

\bibitem{pulsed_yoon2022}
H.~Yoon, P.~Liu, Y.~Park, G.~Choi, P.-P. Choi, H.~Sohn, Pulsed laser-assisted
  additive manufacturing of {Ti-6Al-4V} for in-situ grain refinement,
  Scientific Reports 12~(1) (2022) 22247.

\bibitem{yang2024phase}
C.~Yang, F.~Yang, X.~Meng, S.~N. Putra, M.~Bachmann, M.~Rethmeier, Phase-field
  simulation of the dendrite growth in aluminum alloy {AA5754} during
  alternating current electromagnetic stirring laser beam welding,
  International Journal of Heat and Mass Transfer 218 (2024) 124754.

\bibitem{durga2021grain}
A.~Durga, N.~H. Pettersson, S.~B.~A. Malladi, Z.~Chen, S.~Guo, L.~Nyborg,
  G.~Lindwall, Grain refinement in additively manufactured ferritic stainless
  steel by in situ inoculation using pre-alloyed powder, Scripta Materialia 194
  (2021) 113690.

\bibitem{julien2021nucleation}
I.~Cazic, J.~Zollinger, S.~Mathieu, M.~El~Kandaoui, P.~Plapper, B.~Appolaire,
  New insights into the origin of fine equiaxed microstructures in additively
  manufactured inconel 718, Scripta Materialia 195 (2021) 113740.

\bibitem{samy2023_cet}
V.~P.~N. Samy, M.~Sch{\"a}fle, F.~Brasche, U.~Krupp, C.~Haase, Understanding
  the mechanism of columnar--to-equiaxed transition and grain refinement in
  additively manufactured steel during laser powder bed fusion, Additive
  Manufacturing 73 (2023) 103702.

\bibitem{nabavizadeh2020_cet}
S.~A. Nabavizadeh, M.~Eshraghi, S.~D. Felicelli, Three-dimensional phase field
  modeling of columnar to equiaxed transition in directional solidification of
  {Inconel} 718 alloy, Journal of Crystal Growth 549 (2020) 125879.

\bibitem{Lee2010}
L.~Yuan, P.~D. Lee, Dendritic solidification under natural and forced
  convection in binary alloys: {2D} versus {3D} simulation, Modelling Simul.
  Mater. Sci. Eng. 18~(5) (2010) 055008.

\bibitem{damping_mushy}
W.~Yang, K.-M. Chang, W.~Chen, S.~Mannan, J.~DeBarbadillo, Freckle criteria for
  the upward directional solidification of alloys, Metallurgical and Materials
  Transactions A 32 (2001) 397--406.

\bibitem{yu2022impact}
Y.~Yu, L.~Wang, J.~Zhou, H.~Li, Y.~Li, W.~Yan, F.~Lin, Impact of fluid flow on
  the dendrite growth and the formation of new grains in additive
  manufacturing, Additive Manufacturing 55 (2022) 102832.

\bibitem{tang2022phase}
C.~Tang, H.~Du, Phase field modelling of dendritic solidification under
  additive manufacturing conditions, JOM 74~(8) (2022) 2996--3009.

\bibitem{xiong2022evaluate}
F.~Xiong, Z.~Gan, J.~Chen, Y.~Lian, Evaluate the effect of melt pool convection
  on grain structure of {IN625} in laser melting process using experimentally
  validated process-structure modeling, Journal of Materials Processing
  Technology 303 (2022) 117538.

\bibitem{rai2016ca}
A.~Rai, M.~Markl, C.~K{\"o}rner, A coupled cellular automaton--lattice
  boltzmann model for grain structure simulation during additive manufacturing,
  Computational Materials Science 124 (2016) 37--48.

\bibitem{wu2024unified}
J.~Wu, D.~Sun, W.~Chen, Z.~Chai, A unified lattice {Boltzmann}-phase field
  scheme for simulations of solutal dendrite growth in the presence of melt
  convection, International Journal of Heat and Mass Transfer 220 (2024)
  124958.

\bibitem{kinetic_effects}
H.~Xing, H.~Jing, X.~Dong, L.~Wang, Y.~Han, R.~Hu, Cellular growth during rapid
  directional solidification: Insights from quantitative phase field
  simulations, Materials Today Communications 30 (2022) 103170.

\bibitem{li2023modeling}
C.~Li, J.~Wen, L.~Wang, G.~Lei, Q.~Chen, Modeling of solid-air multi-dendrite
  growth evolution driven by coupled thermal-solute using non-isothermal
  quantitative phase field method, International Communications in Heat and
  Mass Transfer 145 (2023) 106841.

\bibitem{mullis2010antitrapping}
A.~Mullis, J.~Rosam, P.~Jimack, Solute trapping and the effects of
  anti-trapping currents on phase-field models of coupled thermo-solutal
  solidification, Journal of Crystal Growth 312~(11) (2010) 1891--1897.

\bibitem{radhakrishnan2018_3d}
B.~Radhakrishnan, S.~B. Gorti, J.~A. Turner, R.~Acharya, J.~A. Sharon,
  A.~Staroselsky, T.~El-Wardany, Phase field simulations of microstructure
  evolution in {IN718} using a surrogate {Ni-Fe-Nb} alloy during laser powder
  bed fusion, Metals 9~(1) (2018) 14.

\bibitem{nestler2011_opinion}
B.~Nestler, A.~Choudhury, Phase-field modeling of multi-component systems,
  Current opinion in solid state and Materials Science 15~(3) (2011) 93--105.

\bibitem{babu_velocity_pattern}
A.~Plotkowski, M.~M. Kirka, S.~Babu, Verification and validation of a rapid
  heat transfer calculation methodology for transient melt pool solidification
  conditions in powder bed metal additive manufacturing, Additive Manufacturing
  18 (2017) 256--268.

\bibitem{moat2009crystallographic}
R.~Moat, A.~Pinkerton, L.~Li, P.~Withers, M.~Preuss, Crystallographic texture
  and microstructure of pulsed diode laser-deposited {Waspaloy}, Acta
  Materialia 57~(4) (2009) 1220--1229.

\end{thebibliography}
\end{document}